\documentclass[]{jwaa}
\usepackage{graphics}
\begin{document}


\title{The interstellar C$^{18}$O/C$^{17}$O ratio in the solar neighbourhood: 
The $\rho$~Oph cloud\thanks{Based on observations collected with the Swedish/ESO
Submillimeter Telescope (SEST) at the 
European Southern Observatory, Chile (ESO 62.I-0752).
All spectra (some of which are shown in Fig.\ref{spectra}) are available in electronic 
form at the CDS via anonymous ftp to cdsarc.u-strasbg.fr (130.79.128.5)
or via http://cdsweb.u-strasbg.fr/cgi-bin/qcat?J/A+A/
}
}

\author{J.G.A. Wouterloot
	  \inst{1,2} 
 	  \and 
          J. Brand
	  \inst{3}
          \and
         C. Henkel
          \inst{4}
	   }

\offprints{J.G.A. Wouterloot (j.wouterloot@jach.hawaii.edu)}

\institute{Joint Astronomy Centre, 660 N. A'ohoku Place, University Park, 
96720 Hilo, Hawaii, USA \and
Radioastronomisches Institut, Univ. Bonn, Auf dem H\"ugel 71, 
53121 Bonn, Germany  \and
Istituto di Radioastronomia, CNR, Via Gobetti 101, 40129 Bologna, 
Italy \and Max-Planck-Institut f\"ur Radioastronomie,
Auf dem H\"ugel 69, 53121 Bonn, Germany}

\date{Received ;accepted }


\abstract{
Observations of up to ten carbon monoxide (CO and isotopomers) transitions are 
presented to study the interstellar C$^{18}$O/C$^{17}$O ratio towards 21 positions
in the nearby ($d$$\sim$140\,pc) low-mass star forming cloud $\rho$~Oph. 
A map of the C$^{18}$O $J$=1--0 distribution of parts of the cloud is also
shown.
An average $^{12}$C$^{18}$O/$^{12}$C$^{17}$O isotopomeric ratio of 4.11 $\pm$ 0.14, 
reflecting the $^{18}$O/$^{17}$O isotope ratio, 
is derived from Large Velocity Gradient (LVG) calculations. From LTE column
densities we derive a ratio of 4.17$\pm$0.26.
These calculations also show that the kinetic temperature decreases
from about 30~K in the cloud envelope to about 10~K in the cloud cores. This decrease is 
accompanied by an increase of the average molecular hydrogen density
from 10$^4$~cm$^{-3}$ to $\ga$10$^5$~cm$^{-3}$. Towards some lines of sight
C$^{18}$O optical depths reach values of order unity.
}

\maketitle

 %
%

\section{Introduction}

Abundance ratios of interstellar isotopomers are a powerful tool to study 
the chemical evolution of the Galaxy. One such ratio is that of the rare 
species of oxygen, $^{18}$O and $^{17}$O, as measured from the isotopomers 
of CO. For the galactic disk and -center region, Penzias (\cite{penzias}) 
reported average $^{18}$O/$^{17}$O ratios of 3.65$\pm$0.15 and 3.5$\pm$0.2, 
respectively. He also found that the $^{18}$O/$^{17}$O ratio, determined 
from the integrated line intensity ratios 
$\int$T[$^{12}$C$^{18}$O(1$-$0)]d$v$/$\int$T[$^{12}$C$^{17}$O(1$-$0)]d$v$, 
shows no significant gradient with galactocentric distance $R$ out to 10~kpc 
(solar circle: $R_0$=8.5\,kpc):
the $^{18}$O/$^{17}$O ratios of the galactic disk and -center are, within the 
limits of observational accuracy, identical. Models of the chemical evolution
of the Galaxy by Prantzos et al. 
(\cite{prantzos}) suggest that after a few Gyr the ratios in the Galaxy 
should be independent of galactocentric radius. There is, however, a 
discrepancy between the interstellar medium (ISM) values and the much higher
(5.5; Anders \& Grevesse \cite{anders}) solar system one. 
Heikkil\"a et al. 
(\cite{heikkila}) obtain a low value
of 1.6$\pm$0.3 in the LMC, while in the nuclear starbursts NGC\,253 and NGC\,494 5
$^{18}$O/$^{17}$O $\sim$ 6.5 (Harrison et al. \cite{harrison}; Wang et al. \cite{wang}).
These results suggest that  
the $^{18}$O/$^{17}$O ratio depends on metallicity.

The sources observed by Penzias are located in a limited range of galactocentric 
radius (and therefore metal abundance), and we (Wouterloot et al., in preparation) 
have reobserved these sources 
with higher angular resolution and have extended our study to sources out to 
$R$=16\,kpc. While Penzias (\cite{penzias}) only 
observed the $J$=1--0 transition, our observations also include the $J$=2--1 
rotational lines. The goal of the present paper is to study in detail 
excitation and opacity effects that could affect the measured $^{18}$O/$^{17}$O 
ratios and radial gradients on small scales.
These effects are usually either ignored 
or physical parameters are derived by assuming a fixed $^{18}$O/$^{17}$O ratio, so that 
  a more careful study is desirable. 

We have chosen the \object{$\rho$~Oph cloud} ($d$$\sim$140\,pc) 
because of the large range in column densities found therein, and because it is in 
the solar neighbourhood
so that a high linear resolution can be attained towards this object.
Twenty one positions 
were selected for observations from a C$^{18}$O(1--0) map of Wilking \& Lada (\cite{wilking})
to have a range of C$^{18}$O(1--0) intensities.
Towards the positions with the strongest lines we not only 
observed 
the $J$=1--0 and 2--1 lines of four isotopomers ($^{12}$C$^{16}$O, 
hereafter $^{12}$CO; $^{13}$C$^{16}$O, hereafter $^{13}$CO; $^{12}$C$^{18}$O, 
hereafter C$^{18}$O; $^{12}$C$^{17}$O, hereafter C$^{17}$O), but we also measured 
the $J$=3--2 lines of C$^{18}$O and C$^{17}$O.

\section{Observations}

\subsection{SEST observations}

Between September 19 and October 5, 1987 we used the 15-m Swedish-ESO Submillimeter 
Telescope (SEST) to map C$^{18}$O(1--0) towards a part of the \object{$\rho$~Oph 
cloud}. We employed a Schottky receiver in combination with an acousto-optical 
spectrometer (AOS) which had a channel separation of about 43~kHz (0.12~km\,s$^{-1}$). 
All observations were made using frequency switching. The spectra were folded and 
subsequently resampled to a channel width of 0.24~km\,s$^{-1}$. The rms of the resampled 
data is typically 0.14~K ($T_{\rm A}^*$). A 15\arcmin $\times$ 15\arcmin\ region 
was observed with a 40\arcsec\ (in part of the map 20\arcsec) raster (the beam 
size of the SEST at 110~GHz is about 47\arcsec). The mapped region contains the 
cloud cores \object{$\rho$Oph~B1}, \object{$\rho$Oph~C}, \object{$\rho$Oph~E}, and 
\object{$\rho$Oph~F}, as defined from DCO$^+$ maps by Loren et al. (\cite{loren}).

\medskip\noindent
Between January 26 and 31, 1999 we used two SIS receivers at the SEST to observe 
$^{12}$CO, $^{13}$CO, C$^{18}$O, and C$^{17}$O $J$=1--0 and 2--1 towards 21 positions 
in the \object{$\rho$~Oph cloud}. 
The observed transitions and frequencies are given in Cols. 1 and 2 of Table~\ref{freqs}. 
Col. 3 gives the beam size and Cols. 4 and 5 the main beam- and moon efficiencies 
of the telescope used (Col. 6). Most intensities in this paper are 
on the $T_{\rm A}^*$ scale because then uncertain corrections to $T_{\rm R}^*$ 
or $T_{\rm{mb}}$ do not affect the discussed ratios. In some places, i.e. when 
comparing lines from different rotational transitions, the use of a different 
scale (main beam brightness temperature, $T_{\rm{mb}}$, or the average of $T_{\rm{A}}^*$
and $T_{\rm{mb}}$) was unavoidable. This is then mentioned explicitly. 
The pointing accuracy was about 5\arcsec . 

The observed positions were selected from the C$^{18}$O map of Wilking \& Lada 
(\cite{wilking}), and span a large range in C$^{18}$O intensity and hence in $A_v$; 
they are listed in Table~\ref{sources}, in order of decreasing intensity (as 
estimated from the Wilking \& Lada map). Given are a reference number in Col. 1; 
the position in equatorial coordinates in Cols. 2 and 3; the offset positions with 
respect to $\alpha$(1950)=16$^{\rm{h}}$24$^{\rm{m}}$10$^{\rm{s}}$, 
$\delta$(1950)=$-$24\degr23\arcmin\ (this is the average of the H$_2$CO (Martin-Pintado et al. 
\cite{martinpintado}) and NH$_3$ (Zeng et al. \cite{zeng}) positions determined for 
core B1 with the 100-m telescope at 
Effelsberg)[$\alpha$(2000)=16$^{\rm{h}}$27$^{\rm{m}}$11.6$^{\rm{s}}$, 
$\delta$(2000)=$-$24\degr29\arcmin42\arcsec], and the rms in the observed transitions 
in Cols. 6 to 15.

Towards all positions (1 to 21) we observed each of the four isotopomers simultaneously
in the $J$=1--0 and 2--1 transitions. The observations were made using frequency 
switching and we used the high resolution spectrometer (AOS channel-spacing about 
43~kHz) split into two equal parts. Integration times were chosen to obtain similar 
signal to noise ratios in the C$^{17}$O and C$^{18}$O spectra in order to accurately 
derive the line ratios.  In addition to the single positions we made small (3$\times$3) 
maps centered on the selected 21 positions on a 20\arcsec\ raster in C$^{18}$O ($J$=2--1) 
to see whether the positions are located in regions with large gradients where pointing 
errors can influence the observed line ratios, and to be able to convolve the $J $=2--1 data to the 
$J$=1--0 angular resolution.

\begin{figure}
 \resizebox{\hsize}{!}{\includegraphics{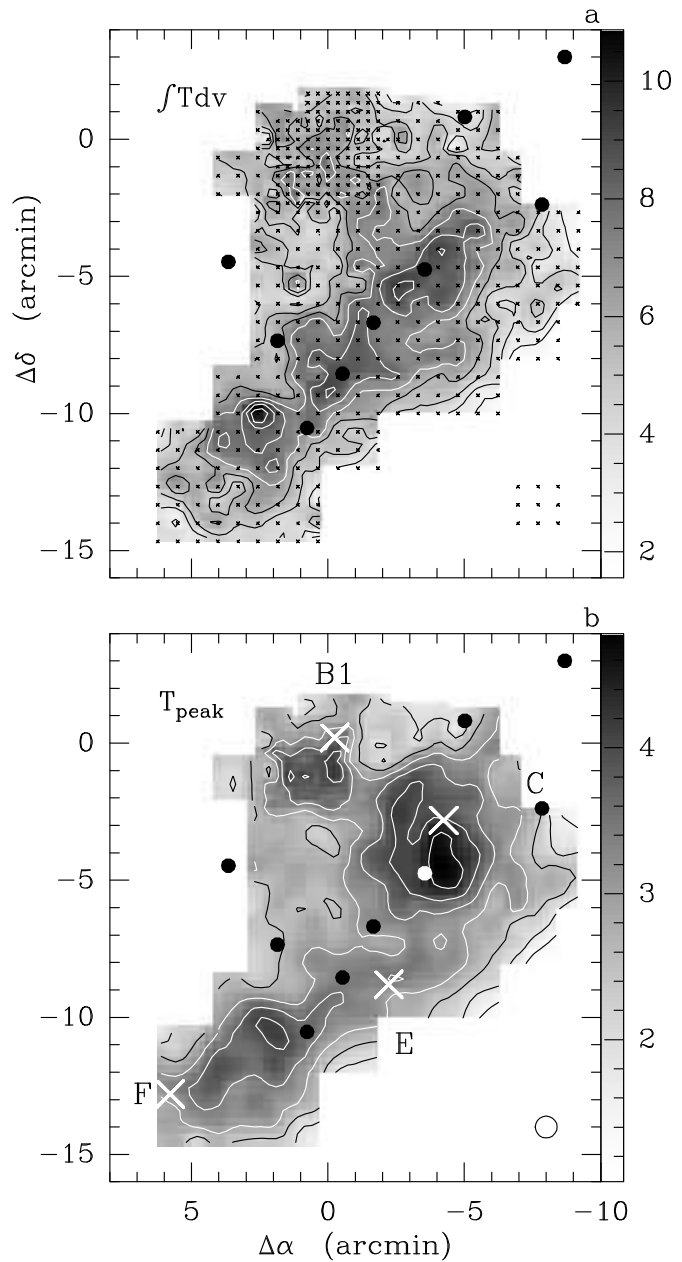}}
 \caption{Distribution of C$^{18}$O(1--0) emission towards the \object{$\rho$~Oph cloud}.
{\bf {a.}} Integrated between $-$1 and +8~km\,s$^{-1}$. Contour levels are 2 to 10~K\,km\,s$^{-1}$ 
in steps of 1~K\,km\,s$^{-1}$. Small crosses indicate the observed positions. Offsets are with 
respect to $\alpha$(1950)=16$^{\rm{h}}$24$^{\rm{m}}$10$^{\rm{s}}$, $\delta$(1950)=$-$24\degr23\arcmin.
The filled circles are those of the 21 positions (see Table 2) that were also observed in other 
isotopomers and that are located within or very near the mapped region. {\bf {b.}} Peak 
$T_{\rm A}^*$ in the same velocity interval. Contour levels are 0.5 to 4.5~K in steps of 
0.5~K. The open circle indicates the angular resolution of the SEST. The crosses indicate 
in order of decreasing declination the DCO$^+$ cores B1, C, E, and F (Loren et al. \cite{loren}).
}
 \label{c18omap}
\end{figure}

\subsection{JCMT observations}

On July 14, 2001, we observed C$^{17}$O(3--2) towards eight of the 21 positions with 
the 15-m James Clerk Maxwell Telescope (JCMT) using frequency switching. The velocity resolution of the 
autocorrelation spectrometer was 0.14~km\,s$^{-1}$ and the rms noise level ranged from 
0.07 to 0.14~K, depending on the line intensity (we tried to reach similar signal-to-noise ratios
at all positions). 

\medskip\noindent
On February 28, 2002 we observed in the 
same way C$^{18}$O(3--2) towards six of the positions previously observed in C$^{17}$O(3--2). 
The rms listed in Cols. 8 and 11 of Table~\ref{sources} was 0.14 to 0.33~K.

\begin{table}
\caption[]{Observed transitions
\label{freqs}}
\begin{flushleft}
\begin{tabular}{lrcccc}
\hline\noalign{\smallskip}
Molecule & Frequency& Beam & $\eta$$_{mb}$ & $\eta$$_{moon}$ & Tel.\\
 & (MHz) &  & & & \\
\hline\noalign{\smallskip}
$^{12}{\rm C}^{18}{\rm O}$(1$-$0)&109782.160 & 47\arcsec & 0.7 & 0.9 & SEST\\
$^{13}{\rm C}^{16}{\rm O}$(1$-$0)&110201.353 & 47\arcsec & 0.7 & 0.9 & SEST\\
$^{12}{\rm C}^{17}{\rm O}$(1$-$0)&112358.988 & 46\arcsec & 0.7 & 0.9 & SEST\\
$^{12}{\rm C}^{16}{\rm O}$(1$-$0)&115271.204 & 45\arcsec & 0.7 & 0.9 & SEST\\
$^{12}{\rm C}^{18}{\rm O}$(2$-$1)&219560.319 & 24\arcsec & 0.5 & 0.9 & SEST\\
 $^{13}{\rm C}^{16}{\rm O}$(2$-$1)&220398.686 & 24\arcsec & 0.5 & 0.9 & SEST\\
$^{12}{\rm C}^{17}{\rm O}$(2$-$1)&224714.368 & 24\arcsec & 0.5 & 0.9 & SEST\\
$^{12}{\rm C}^{16}{\rm O}$(2$-$1)&230537.990 & 23\arcsec & 0.5 & 0.9 & SEST\\
$^{12}{\rm C}^{18}{\rm O}$(3$-$2)&329330.545 & 14\arcsec & 0.6 & 0.9 & JCMT\\
$^{12}{\rm C}^{17}{\rm O}$(3$-$2)&337061.130 & 14\arcsec & 0.6 & 0.9 & JCMT\\
\noalign{\smallskip}
\hline
\noalign{\smallskip}
\end{tabular}
\end{flushleft}
\end{table}
%
%

%
%
\begin{table*}
\caption[]{Observed sources and rms values (K on a $T_{\rm A}^*$ scale) for 
individual transitions.
\label{sources}}
\begin{flushleft}
\begin{tabular}{lccrrrrrrrrrrrr}
\hline\noalign{\smallskip}
\multicolumn{1}{l}{Pos.}& \multicolumn{1}{c}{$\alpha$(1950)}& 
\multicolumn{1}{c}{$\delta$(1950)}& 
\multicolumn{2}{c}{Offset$^{a}$}&
\multicolumn{3}{c}{rms C$^{17}$O}&
\multicolumn{3}{c}{rms C$^{18}$O}&
\multicolumn{2}{c}{rms $^{13}$CO}&
\multicolumn{2}{c}{rms $^{12}$CO}\\
\multicolumn{1}{l}{}& \multicolumn{1}{c}{{\sl h\ m\ s}}&
\multicolumn{1}{c}{\degr\ \arcmin\ \arcsec}& 
\multicolumn{2}{c}{\arcmin}&
\multicolumn{1}{c}{1--0}&
\multicolumn{1}{c}{2--1}&
\multicolumn{1}{c}{3--2}&
\multicolumn{1}{c}{1--0}&
\multicolumn{1}{c}{2--1}&
\multicolumn{1}{c}{3--2}&
\multicolumn{1}{c}{1--0}&
\multicolumn{1}{c}{2--1}&
\multicolumn{1}{c}{1--0}&
\multicolumn{1}{c}{2--1}\\
\hline\noalign{\smallskip}
1 & 16 23 54.0 & -24 27 45 & -3.55 & -4.75 & 0.023& 0.029& 0.085 &0.050& 0.070& 0.14 & 0.12& 0.20& 0.20& 0.23\\
2 & 16 23 11.6 & -24 14 20 & -13.20 & 8.27 & 0.025& 0.042& 0.140  &0.068& 0.091& 0.21 & 0.07& 0.14& 0.27& 0.26\\
3 & 16 24 02.1 & -24 31 33 & -0.53  & -8.55 & 0.018& 0.032& 0.071 &0.070& 0.066& 0.19 & 0.15& 0.20& 0.30& 0.30\\
4 & 16 24 02.7 & -24 29 41 & -1.67 & -6.68 & 0.019& 0.025& 0.081&0.053& 0.078& 0.20 & 0.15& 0.22& 0.27& 0.26\\
5 & 16 23 28.6 & -24 16 34 & -9.43 & 6.60 & 0.017& 0.031& 0.076&0.052& 0.077& 
0.33 & 0.13& 0.20& 0.30& 0.30\\
6 & 16 24 13.4 & -24 33 32 & 0.77 & -10.53 & 0.026& 0.028& 0.086&0.060& 0.087& 0.28 & 0.17& 0.20& 0.30& 0.33\\
7 & 16 23 31.9 & -24 20 00 & -8.68 & 3.00 & 0.016& 0.029& 0.145&0.058& 0.077& &0.14& 0.21& 0.34& 0.30\\
8 & 16 24 18.1 & -24 30 21 & 1.85 & -7.35 & 0.013& 0.020& 0.103&0.045& 0.051& &0.12& 0.19& 0.35& 0.33\\
9 & 16 23 11.2 & -24 13 30 & -13.38 & 9.50 & 0.020& 0.029& &0.072& 0.074& &0.12& 0.20& 0.34& 0.30\\
10 & 16 23 48.0 & -24 22 11 & -5.02 & 0.82 & 0.019& 0.031& &0.104& 0.063& &0.14& 0.19& 0.39& 0.34\\
11 & 16 24 26.0 & -24 27 28 & 3.65 & -4.47 & 0.023& 0.033& &0.079& 0.059& &0.15& 0.18& 0.33& 0.28\\
12 & 16 23 33.8 & -24 10 48 & -8.25 & 12.20 & 0.018& 0.025& &0.071& 0.059& &0.14& 0.22& 0.36& 0.33\\
13 & 16 23 35.5 & -24 25 23 & -7.85 & -2.38 & 0.012& 0.020& &0.067& 0.060& &0.13& 0.24& 0.38& 0.33\\
14 & 16 23 50.0 & -24 18 20 & -4.55 & 4.67 & 0.016& 0.026& &0.071& 0.049& &0.16& 0.20& 0.39& 0.36\\
15 & 16 23 18.3 & -24 18 14 & -11.77 & 4.77 & 0.012& 0.017& &0.076& 0.051& &0.16& 0.23& 0.36& 0.34\\
16 & 16 23 20.5 & -24 23 06 & -11.27 & -0.10 & 0.0072& 0.014& &0.031& 0.030& &0.15& 0.21& 0.38& 0.35\\
17 & 16 24 20.9 & -24 15 23 & 2.48 & 7.62 & 0.011& 0.016& &0.057& 0.042& &0.13& 0.22& 0.33& 0.34\\
18 & 16 23 12.8 & -24 09 16 & -13.02 & 13.73 & 0.011& 0.012& &0.051& 0.041& &0.13& 0.22& 0.32& 0.28\\
19 & 16 23 05.8 & -24 23 06 & -14.62 & -0.10 & 0.010& 0.019& &0.038& 0.032& &0.16& 0.21& 0.35& 0.32\\
20 & 16 24 33.4 & -24 15 23 & 5.33 & 7.62 &  0.0066& 0.010& &0.034& 0.037& &0.13& 0.19& 0.32& 0.38\\
21 & 16 23 25.6 & -24 07 07 &  -10.12 & 15.88 & 0.010& 0.017& &0.042& 0.050& &0.14& 0.23& 0.28& 0.30\\
\noalign{\smallskip}
\hline
\multicolumn{14}{l}{{\it a}. With respect to $\alpha$(1950)=16$^{\rm{h}}$24$^{\rm{m}}$10$^{\rm{s}}$, 
                     $\delta$(1950)=$-$24\degr23\arcmin}\\
\noalign{\smallskip}
\end{tabular}
\end{flushleft}
\end{table*}
%
%

\section{Results}

\begin{figure}
 \resizebox{\hsize}{!}{\includegraphics{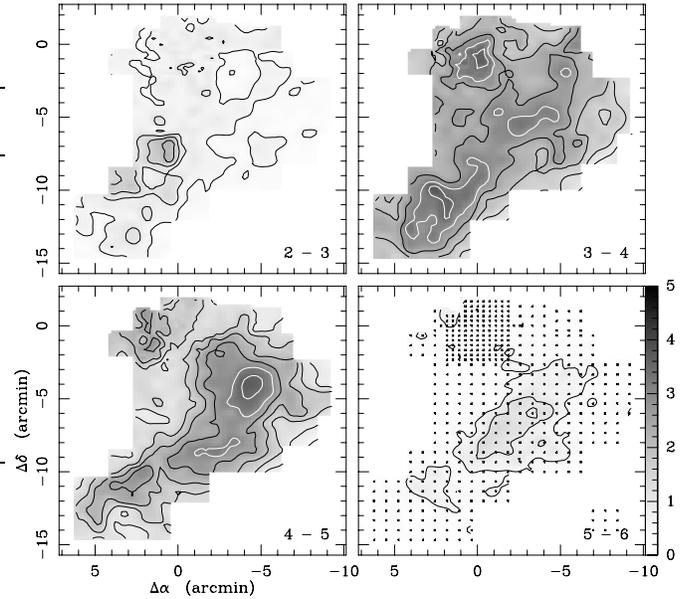}}
 \caption{C$^{18}$O(1--0) distribution in four velocity intervals (given in the lower right 
hand corner of each panel). Contour levels are 0.5 to 4.5~K\,km\,s$^{-1}$ in steps of 
0.5~K\,km\,s$^{-1}$.
}
 \label{rophb_chan}
\end{figure}

The C$^{18}$O(1--0) distribution in the mapped region is shown in Fig.\,\ref{c18omap}. 
The map includes the cloud cores \object{$\rho$Oph~B1}, \object{$\rho$Oph~C}, 
\object{$\rho$Oph~E}, and \object{$\rho$Oph~F} (see e.g. the 1.3~mm continuum map in
Fig.\,1 of Motte et al. (\cite{motte}) which indicates the locations of the cores A, B1, 
B2, C, D, E, and F). These cores were originally defined in DCO$^+$ maps by Loren et al. 
(\cite{loren}). We show the emission integrated over all velocities (--1 to 
8~km\,s$^{-1}$) in Fig.\,\ref{c18omap}a and the peak $T_{\rm A}^*$ distribution  
in Fig.\,\ref{c18omap}b, respectively. A comparison between the panels (and Gaussian 
fits to the lines) shows that \object{$\rho$Oph~B1} and \object{$\rho$Oph~C} (near 
offsets (0, 0) and (--4,--3) respectively) have relatively narrow lines (about 
1.5~km\,s$^{-1}$) and a (more) pronounced peak in Fig.\,\ref{c18omap}b, whereas broader 
(about 2.3~km\,s$^{-1}$) lines occur north of \object{$\rho$Oph~E}, near 
($-1,-8$). At 
the edges of the map, near \object{$\rho$Oph~F}, southwest of \object{$\rho$Oph~C}, 
and in between \object{$\rho$Oph~B1} and \object{$\rho$Oph~C} line widths reach  values 
of about 1.0~km\,s$^{-1}$. Compared to the lower angular resolution C$^{18}$O(1--0) map 
in Fig.\,2 of Wilking \& Lada (\cite{wilking}; beam size 1.1\arcmin\ on a 1\arcmin\ or 
2\arcmin\ raster), the distribution in Fig.\,\ref{c18omap} shows finer spatial structure. 
There is also a reasonable correlation between the 
1.3\,mm continuum in Fig.\,1 of Motte et al. (\cite{motte}) and the C$^{18}$O distribution. 
The C$^{18}$O $J$=1--0 emission in four 1~km\,s$^{-1}$-wide channels is shown in 
Fig.\,\ref{rophb_chan}. \object{$\rho$Oph~C} and \object{$\rho$Oph~E} are mainly observed 
at 4$-$5~km\,s$^{-1}$, whereas \object{$\rho$Oph~B1} and \object{$\rho$Oph~F} show most 
emission at 3$-$4~km\,s$^{-1}$.

All spectra measured towards the first three of the twenty-one positions (Table~\ref{sources}) are shown 
in Fig.\,\ref{spectra} (the spectra towards all positions are published in the Appendix in the 
electronic edition).
The velocity interval is $-$8 to +16~km\,s$^{-1}$. Most $^{12}$CO 
and some $^{13}$CO spectra show self-absorption (a clear minimum in between two peaks 
in the spectra that is not seen in the lines of the rarer CO isotopomers).
If flat-topped spectra are also considered as an indication for 
self-absorption, this phenomenon occurs at even more positions. The numbers in the boxes 
of the C$^{18}$O(2--1) profiles show the values of $\int$$T_{\rm A}^*$[C$^{18}$O(2--1)]d$v$ 
for the 20\arcsec -spaced nine-point map around each position. Some positions show 
significant (10 to 15\%) intensity gradients where pointing differences between C$^{18}$O 
and C$^{17}$O (which were not observed simultaneously) could influence the derived line 
ratios. Equivalent line widths ($\int$$T_{\rm A}^*$d$v$/1.06$T_{\rm A}^*$[peak]) for 
 C$^{18}$O(1--0) range between 0.84~km\,s$^{-1}$ (pos. 14) and 1.91~km\,s$^{-1}$ (pos. 8). 
At many positions the C$^{18}$O and C$^{17}$O line profiles show the presence of several 
velocity components, which are more pronounced in $^{13}$CO and $^{12}$CO (but at slightly 
different velocities, probably due to self-absorption and saturation in the more abundant 
isotopomers). Two of these velocity components are visible (at 3--4 and 4--5~km\,s$^{-1}$)
in the C$^{18}$O(1--0) channel maps of Fig.\,\ref{rophb_chan}. The C$^{17}$O spectra also 
show hyperfine structure (see e.g. Lovas \& Krupenie \cite{lovas}).

\begin{figure*}
 \resizebox{17cm}{!}{\includegraphics{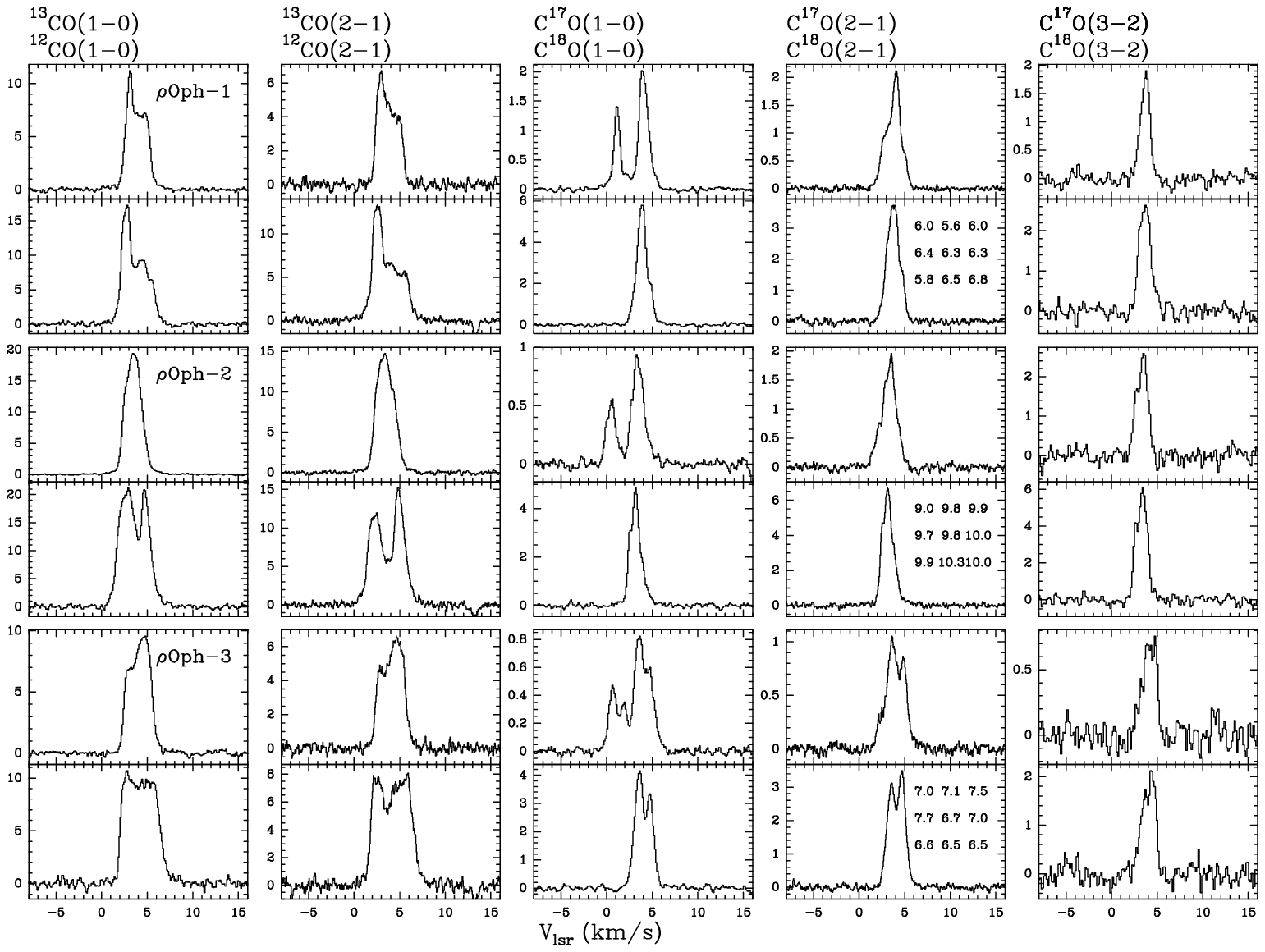}}
 \caption{
$^{12}$CO, $^{13}$CO, C$^{18}$O and C$^{17}$O spectra towards the first three of the selected 
positions given in Table 2. Numbers in the C$^{18}$O(2--1) panels indicate integrated line intensities in 
                      K\,km\,s$^{-1}$ on a 20\arcsec\ raster centered at the respective position. The complete set of spectra 
can be obtained in FITS format at http://cdsweb.u-strasbg.fr/cgi-bin/qcat?J/A+A/
}
 \label{spectra}
\end{figure*}

\noindent

%
%
\begin{table*}
\caption[]{Observed C$^{17}$O, C$^{18}$O, and $^{13}$CO  (1--0), (2--1), and 
(3--2) integrated line intensities
and line ratios (i.e. not corrected for the frequency difference). For each position
the second line gives the formal errors not including calibration uncertainties.
\label{results}}
\begin{flushleft}
 \begin{tabular}{rccccccccccccc}
\hline\noalign{\smallskip}
Pos. &
\multicolumn{3}{c}{C$^{17}$O}&
\multicolumn{3}{c}{C$^{18}$O} &
\multicolumn{2}{c}{$^{13}$CO} &
\multicolumn{1}{c}{${\rm{C^{18}O}}\over{\rm{C^{17}O}}$} &
\multicolumn{1}{c}{${\rm{C^{18}O}}\over{\rm{C^{17}O}}$} &
\multicolumn{1}{c}{${\rm{C^{18}O}}\over{\rm{C^{17}O}}$} &
\multicolumn{1}{c}{${\rm{^{13}CO}}\over{\rm{C^{18}O}}$} &
\multicolumn{1}{c}{${\rm{^{13}CO}}\over{\rm{C^{18}O}}$} \\
 &
\multicolumn{1}{c}{(1--0)} &
\multicolumn{1}{c}{(2--1)} &
\multicolumn{1}{c}{(3--2)} &
\multicolumn{1}{c}{(1--0)} &
\multicolumn{1}{c}{(2--1)} &
\multicolumn
{1}{c}{(3--2)} &
\multicolumn{1}{c}{(1--0)} &
\multicolumn{1}{c}{(2--1)} &
\multicolumn{1}{c}{(1--0)} &
\multicolumn{1}{c}{(2--1)} &
\multicolumn{1}{c}{(3--2)} &
\multicolumn{1}{c}{(1--0)} &
\multicolumn{1}{c}{(2--1)} \\
 & \multicolumn{8}{c}{K\,km\,s$^{-1}$} & & \\
\hline\noalign{\smallskip}
 1  & 4.12  & 3.79 & 2.03 & 8.59  & 6.60  & 4.06 &24.45  &15.46  & 2.08  & 1.74  & 2.00 & 2.85  & 2.34\\
    & 0.02  & 0.02 & 0.10 & 0.05  & 0.05  & 0.16 & 0.12  & 0.15  & 0.02  & 0.02  & 0.12 & 0.02  & 0.03\\
 2  & 2.25  & 3.52 & 3.52 & 6.82  &10.07  & 8.42 &43.07  &35.38  & 3.03  & 2.86  & 2.39 & 6.31  & 3.51\\
     & 0.03  & 0.03 & 0.16 & 0.07  & 0.07  & 0.24 & 0.08  & 0.10  & 0.05  & 0.03  & 0.13 & 0.07  & 0.03\\
 3  & 2.50  & 2.43 & 1.22 & 8.33  & 7.17  & 3.13 &27.03  &19.12  & 3.33  & 2.95  & 2.57 & 3.24  & 2.66\\
    & 0.02  & 0.02 & 0.08 & 0.07  & 0.05  & 0.22 & 0.15  & 0.15  & 0.04  & 0.03  & 0.25 & 0.03  & 0.03\\
 4  & 2.62  & 2.29 & 1.49 & 7.52  & 6.62  & 3.65 &25.33  &18.31  & 2.87  & 2.89  & 2.45 & 3.37  & 2.77\\
    & 0.02  & 0.02 & 0.09 & 0.05  & 0.06  & 0.23 & 0.16  & 0.16  & 0.03  & 0.03  & 0.22 & 0.03  & 0.03\\
 5  & 1.27  & 2.69 & 3.38 & 4.27  & 6.74  & 8.48 &34.31  &28.68  & 3.36  & 2.50  & 2.51 & 8.03  & 4.26\\
    & 0.02  & 0.02 & 0.09 & 0.05  & 0.06  & 0.38 & 0.14  & 0.15  & 0.06  & 0.03  & 0.13 & 0.11  & 0.04\\
 6  & 2.00  & 2.06 & 0.95 & 7.02  & 6.00  & 3.21 &22.31  &16.41  & 3.51  & 2.91  & 3.37 & 3.18  & 2.73\\
    & 0.03  & 0.02 & 0.10 & 0.06  & 0.06  & 0.33 & 0.18  & 0.15  & 0.06  & 0.04  & 0.49 & 0.04  & 0.04\\
 7  & 1.97  & 3.71 & 4.85 & 6.31  & 8.84  & &33.63  &23.46  & 3.20  & 2.38  & & 5.33  & 2.65\\
    & 0.02  & 0.02 & 0.17 & 0.06  & 0.06  & & 0.14  & 0.15  & 0.04  & 0.02  & & 0.06  & 0.02\\
 8  & 1.47  & 1.37 & 0.57 & 5.20  & 4.84  & &24.02  &15.08  & 3.55  & 3.53  & & 4.61  & 3.12\\
    & 0.01  & 0.01 & 0.12 & 0.05  & 0.04  & & 0.13  & 0.14  & 0.05  & 0.05  & & 0.05  & 0.04\\
 9  & 2.04  & 3.44 & & 7.79  &10.85  & &40.61  &29.41  & 3.82  & 3.15  & & 5.21  & 2.71\\
    & 0.02  & 0.02 & & 0.08  & 0.05  & & 0.12  & 0.15  & 0.05  & 0.02  & & 0.05  & 0.02\\
 10 & 1.41  & 1.63 & & 2.97  & 4.27  & &21.81  &14.73  & 2.10  & 2.62  & & 7.35  & 3.45\\
    & 0.02  & 0.02 & & 0.11  & 0.05  & & 0.14  & 0.14  & 0.08  & 0.05  & & 0.27  & 0.05\\
 11 & 1.25  & 1.17 & & 4.53  & 3.80  & &20.52  &13.59  & 3.62  & 3.24  & & 4.53  & 3.58\\
    & 0.02  & 0.02 & & 0.08  & 0.04  & & 0.15  & 0.13  & 0.09  & 0.08  & & 0.09  & 0.05\\
 12 & 1.20  & 2.36 & & 4.32  & 5.47  & &24.53  &16.41  & 3.60  & 2.32  & & 5.69  & 3.00\\
    & 0.02  & 0.02 & & 0.07  & 0.04  & & 0.14  & 0.16  & 0.08  & 0.03  & & 0.10  & 0.04\\
 13 & 0.83  & 1.38 & & 2.40  & 3.47  & &25.52  &16.69  & 2.88  & 2.52  & &10.63  & 4.80\\
    & 0.01  & 0.01 & & 0.07  & 0.04  & & 0.13  & 0.18  & 0.09  & 0.04  & & 0.31  & 0.08\\
 14 & 0.95  & 1.35 & & 2.87  & 3.30  & &16.25  &14.59  & 3.00  & 2.45  & & 5.66  & 4.43\\
    & 0.02  & 0.02 & & 0.07  & 0.04  & & 0.17  & 0.15  & 0.09  & 0.04  & & 0.16  & 0.07\\
 15 & 1.07  & 2.06 & & 3.71  & 6.40  & &37.64  &33.15  & 3.47  & 3.11  & &10.14  & 5.18\\
    & 0.01  & 0.01 & & 0.08  & 0.04  & & 0.17  & 0.17  & 0.08  & 0.03  & & 0.22  & 0.04\\
 16 & 0.33  & 0.59 & & 1.11  & 1.69  & &19.45  &15.78  & 3.36  & 2.86  & &17.55  & 9.35\\
    & 0.01  & 0.01 & & 0.03  & 0.02  & & 0.16  & 0.15  & 0.12  & 0.06  & & 0.53  & 0.15\\
 17 & 0.65  & 0.67 & & 2.36  & 2.56  & &16.94  &11.33  & 3.65  & 3.80  & & 7.17  & 4.43\\
    & 0.01  & 0.01 & & 0.06  & 0.03  & & 0.13  & 0.16  & 0.11  & 0.08  & & 0.19  & 0.08\\
 18 & 0.55  & 1.05 & & 1.94  & 3.32  & &24.13  &17.47  & 3.55  & 3.15  & &12.46  & 5.27\\
    & 0.01  & 0.01 & & 0.05  & 0.03  & & 0.13  & 0.16  & 0.12  & 0.04  & & 0.35  & 0.07\\
 19 & 0.32  & 0.67 & & 0.98  & 1.62  & &18.85  &15.80  & 3.08  & 2.43  & &19.20  & 9.77\\
    & 0.01  & 0.01 & & 0.04  & 0.02  & & 0.16  & 0.15  & 0.16  & 0.06  & & 0.80  & 0.17\\
 20 & 0.37  & 0.44 & & 1.56  & 1.78  & &17.10  &11.87  & 4.20  & 4.07  & &10.95  & 6.65\\
    & 0.01  & 0.01 & & 0.03  & 0.03  & & 0.14  & 0.14  & 0.12  & 0.09  & & 0.26  & 0.13\\
 21 & 0.40  & 0.88 & & 1.67  & 2.29  & &17.92  &12.85  & 4.16  & 2.61  & &10.71  & 5.61\\
    & 0.01  & 0.01 & & 0.04  & 0.04  & & 0.15  & 0.17  & 0.15  & 0.06  & & 0.30  & 0.12\\
\noalign{\smallskip}
\hline
\end{tabular}
\end{flushleft}
\end{table*}

\bigskip\noindent

\section{Isotopomeric ratios}

\subsection{Isotopomeric ratios derived from line intensities}

Integrated line intensities over the velocity interval $-$1.5 to 8~km\,s$^{-1}$ (i.e. 
over all velocity components) and their ratios at the measured 21 positions are listed in 
Table~\ref{results} for C$^{17}$O, C$^{18}$O, and $^{13}$CO. For each position the 
first line gives the intensities (Cols. 2 to 9) and ratios (Cols. 10 to 14) 
and the second line the uncertainty therein, obtained from the rms in the baseline. 
We note that while we obtained the positions from the C$^{18}$O(1--0) map of Wilking 
\& Lada (\cite{wilking}) and ordered them according to an expected decrease in integrated 
intensity, we see this in our results, but there is significant scatter. This 
is likely caused by a lower angular resolution and undersampling in the Wilking \& Lada map, 
compared to our data. The selected positions cover a large range 
in C$^{18}$O(1--0) integrated intensity (from 0.98 to 8.59~K\,km\,s$^{-1}$) (or $A_{\rm v}$ 
$\approx$ 10 to 200 mag (using column densities in Table~\ref{nratios} and Frerking
et al. \cite{frerking})), as intended. 

\begin{figure}
 \resizebox{\hsize}{!}{\includegraphics{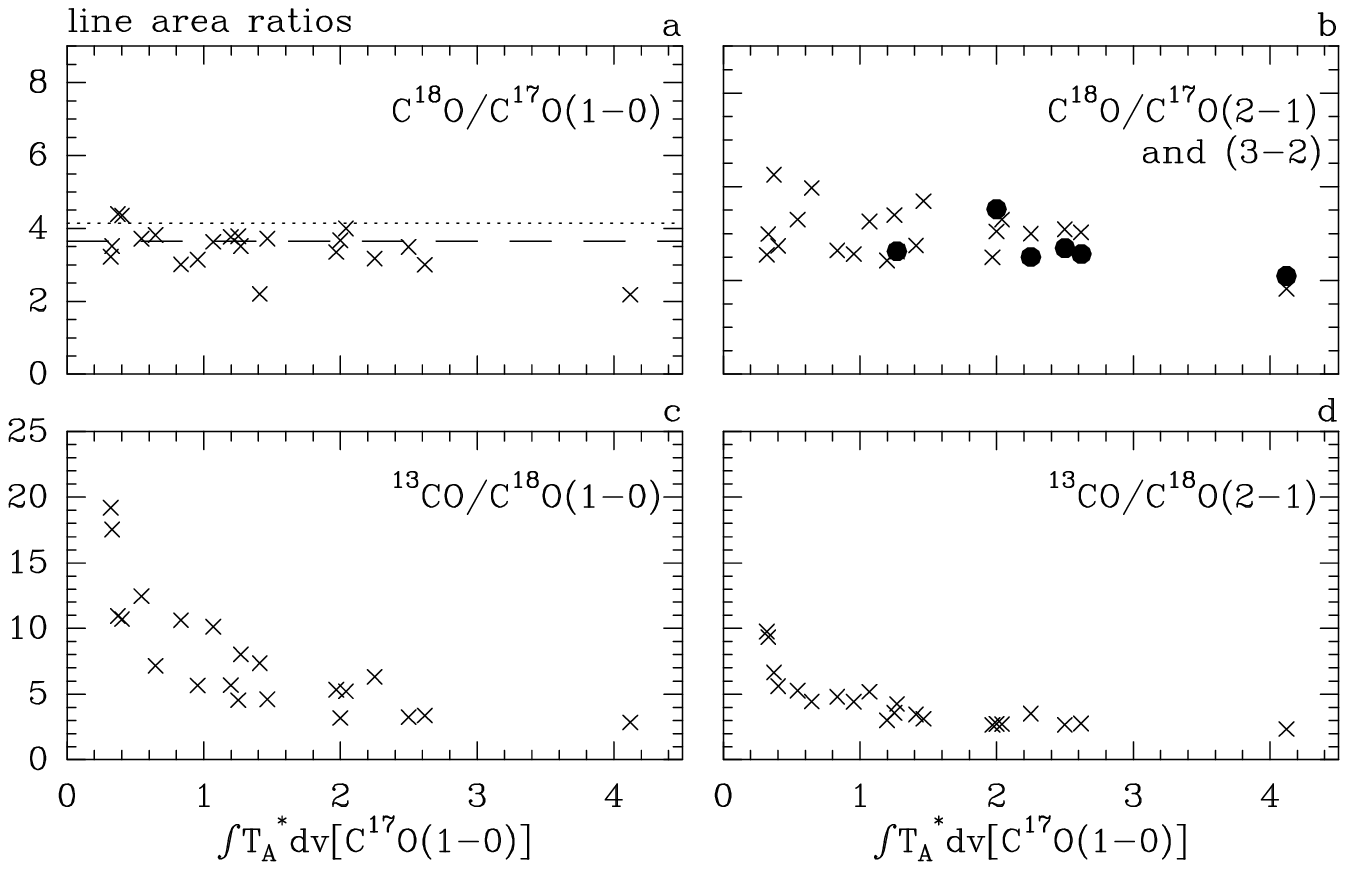}}
 \caption{C$^{18}$O/C$^{17}$O and $^{13}$CO/C$^{18}$O isotopomeric ratios as a function of 
$\int$$T_{\rm A}^*$[C$^{17}$O(1--0)]d$v$ for ({\bf {a,c}}) $J$=1--0, ({\bf {b,d}}) 
$J$=2--1 (crosses) and ({\bf b}) $J$=3--2 (filled circles). The C$^{18}$O/C$^{17}$O 
and $^{13}$CO/C$^{18}$O line area ratios have been corrected for the frequency difference 
(see Sect.\,4.1).
The dashed line indicates the ratio found by Penzias (\cite{penzias}) for the galactic disk, 
and the dotted line denotes the result from Bensch et al. (\cite{bensch}) from $^{13}$C$^{18}$O and 
$^{13}$C$^{17}$O(1--0). }
 \label{coratios}
\end{figure}

The resulting C$^{18}$O/C$^{17}$O and $^{13}$CO/C$^{18}$O integrated line intensity 
ratios for the (1--0) and (2--1) transitions as a function of $\int$$T_{\rm A}^*$[C$^{17}$O(1--0)]d$v$ 
(the most optically thin transition) are shown in Figs.\,\ref{coratios}a-d. Here 
(not in Table~\ref{results}, following Penzias \cite{penzias}) we corrected the ratios 
for the difference in frequency, which amounts for C$^{18}$O/C$^{17}$O to a factor 1.047 
(C$^{18}$O/C$^{17}$O(corrected) 
= ($\nu_{17}/\nu_{18})^2$ C$^{18}$O/C$^{17}$O(observed), see Linke et al. \cite{linke}), 
but not for optical depth and excitation effects. Therefore, if these effects are negligible, 
the corrected ratios should be equivalent to those of the column densities. Formal errors 
 derived from the uncertainties in the line areas (see Table~\ref{results}) are small with 
respect to the errors introduced by the calibration uncertainties. The latter may amount 
to 7\% causing an error in the ratios of 10\%. The 3$\times$3 C$^{18}$O(2--1) maps 
around the observed positions (see Fig.\,\ref{spectra}) show that over an angular scale 
of  20\arcsec\ the change in integrated intensity is typically 10\% or less. This implies 
that pointing errors of 5\arcsec\ cause errors of a few percent in the observed line
intensity ratios. In Fig.\,\ref{coratios}b we also show (as filled circles) the 
C$^{18}$O/C$^{17}$O ratios derived from the JCMT $J$=3--2 observations. These agree well 
with the $J$=2--1 and $J$=1--0 results. For comparison we also show in Fig.\,\ref{coratios}a
the average result from Penzias 
(\cite{penzias}) for the galactic disk as a dashed line, and the value towards our 
pos. 1 derived by Bensch et al. (\cite{bensch}) from $^{13}$C$^{18}$O and 
$^{13}$C$^{17}$O(1--0) observations as a dotted line.

At pos. 1, where integrated C$^{17}$O line intensities and CO column densities are 
highest, the observed C$^{18}$O/C$^{17}$O ratios are significantly lower than towards 
the other positions (see Figs.\,\ref{coratios}a,b). This holds for all three observed 
rotational transitions. At pos. 10 the C$^{18}$O/C$^{17}$O ratio is lower than at 
other positions (except pos. 1) for $J$=1--0 but not for $J$=2--1. Omitting pos. 
1, the unweighted average of the C$^{18}$O/C$^{17}$O 
integrated intensity ratios (including frequency correction)
is 3.53$\pm$0.48 (sd; me 0.11) for the (1--0) and 3.06$\pm$0.49 
(sd; me 0.11) for the (2--1) transition (for average values we derive both the standard 
deviation (sd) and the error in the mean (me=sd/$\sqrt N$), which is the most relevant 
parameter describing the uncertainty of the average values). For $J$=3--2 (positions 
2 to 6) we obtain a ratio of 2.78$\pm$0.40 (sd; me 0.18); for the same positions the 
$J$=1--0 and 2--1 ratios are 3.37$\pm$0.26 (sd; me 0.12) and 2.95$\pm$0.18 (sd; me 0.08), 
respectively, suggesting a decrease of the ratio with $J$, possibly because of increasing 
optical depth (see Sect.\,4.2.3).
At pos. 1 the ratios are 2.18$\pm$0.02 (1--0), 1.82$\pm$0.02 (2--1), 
and 2.09$\pm$0.12 (3--2). 

In Figs.\,\ref{coratios}c,d $^{13}$CO/C$^{18}$O ratios are shown for the (1--0) and (2--1) 
transitions. This ratio may be strongly affected by $^{13}$C fractionation for small column 
densities (Bally \& Langer \cite{bally}, Langer et al. \cite{langergr}), self-shielding 
(van Dishoeck \& Black \cite{dishoeck}), and by high $^{13}$CO optical depths at large column 
densities, which explains the decrease from 10 to 20 in the outer parts of the $ \rho$~Oph
cloud to about 3 in the cloud centre. This decline in integrated line intensity ratios
is similar to that seen in Barnard~5 (Langer et al. (\cite{langerwil}), their Fig.\,5), 
but in $\rho$~Oph $^{13}$CO/C$^{18}$O ratios reach even lower values than in Barnard~5. 
Towards the very outer parts of Barnard~5 where $\int$T[$^{13}$CO(1--0)]d$v$ is 1-4~K\,km\,s$^{-1}$, 
this ratio also decreases. In $\rho$ Oph we do not see this effect, possibly because here 
the $^{13}$CO emission is much stronger towards all observed positions. We note that 
Zielinsky et al. (\cite{zielinsky}) could explain increasing $^{13}$CO/C$^{18}$O ratios 
towards the edge of a Photon Dominated Region (PDR) by the presence of few big clumps in the 
cloud center and many small clumps at the cloud edge.

\subsection{Isotopomeric ratios derived from LTE column densities}

The isotopomeric ratios are 
likely to be affected by optical depth effects, which will be strong for CO (most positions 
show self-absorption), significant for $^{13}$CO, and not negligible for C$^{18}$O. For 
C$^{17}$O small optical depths are expected towards all observed lines of sight. At pos. 
1, which shows the highest C$^{17}$O column density, the observed integrated intensity ratios 
for $J$=1--0 C$^{18}$O/$^{13}$C$^{18}$O and C$^{17}$O/$^{13}$C$^{17}$O are 35.8 and 68.7, 
respectively (Bensch et al. \cite{bensch}), indicating 
(assuming that the $^{12}$C/$^{13}$C ratio is about 70; Wilson \& Rood \cite{wilson}) that 
the C$^{18}$O optical depth 
is about 1.5 whereas the C$^{17}$O optical depth is small. 

Trying to account for optical depth effects, we have derived Local Thermodynamical Equilibrium
(LTE) column densities of $^{13}$CO, C$^{18}$O, and C$^{17}$O. Below we describe how we 
calculate excitation temperatures and optical depths. 

\subsubsection{Excitation temperature}

For the calculation of $T_{\rm {ex}}$, both 
the $T_{\rm A}^*$ and the $T_{\rm mb}^*$ temperatures scales are relevant.
The $T_{\rm mb}$-scale is 
strictly valid in the extreme case of a source only covering the main beam, while the 
$T_{\rm A}^*$-value applies to the opposite extreme, a very extended source. We
have estimated the CO excitation temperatures in two different ways, in each of which
we used both $T_{\rm A}^*$ and $T_{\rm mb}^*$.

Firstly, we derive excitation temperatures from the peak temperatures of CO(1--0) and 
CO(2--1) using
$$T_{\rm {ex}} = 5.532\, \left[ \ln \left(1 + {5.532 \over
(T_{10} + 0.819)} \right) \right]^{-1}, $$
where $T_{10}$ is the peak $T$[$^{12}$CO(1--0)] temperature,
or
$$T_{\rm {ex}} = 11.064\, \left[ \ln \left(1 + {11.064 \over
(T_{21} + 0.187)} \right) \right]^{-1}, $$
where $T_{21}$ is the peak $T$[$^{12}$CO(2--1)] temperature ({\bf method 1}). 
This could underestimate the excitation temperature whenever there is self-absorption. 
The method helps to constrain excitation temperatures, in 
particular for the outer parts of the cloud. 

Secondly we obtain excitation temperatures from the (2--1)/(1--0) ratio of the integrated 
intensities of C$^{18}$O and C$^{17}$O, respectively ({\bf method 2}), 
assuming that the transitions are optically thin 
(if this is not the case (mainly for C$^{18}$O) the excitation temperature will be underestimated) 
and that beam filling effects do not affect the ratio [C$^{17}$O: (2--1)/(1--0) = 
4.0~exp(--10.78/$T_{\rm {ex}}$); C$^{18}$O: (2--1)/(1--0) = 4.0~exp($-$10.54/$T_{\rm {ex}}$)].
The $J$=2--1 C$^{18}$O and C$^{17}$O data were convolved to the $J$=1--0 resolution using the 
nine point C$^{18}$O 
$J$=2--1 maps described in Sect.\,2 (see also Fig.\,3). The corrections are small ($\le$5\% at
most positions, but 16\% at pos. 19).
In contrast to method 1 that uses optically thick CO lines potentially tracing
predominantly cloud envelopes, method 2 is based on tracers that are more representative of
the entire molecular column density. A comparison of excitation temperatures derived by methods
1 and 2 can show whether there are temperature gradients in the cloud.

\begin{figure}
 \resizebox{\hsize}{!}{\includegraphics{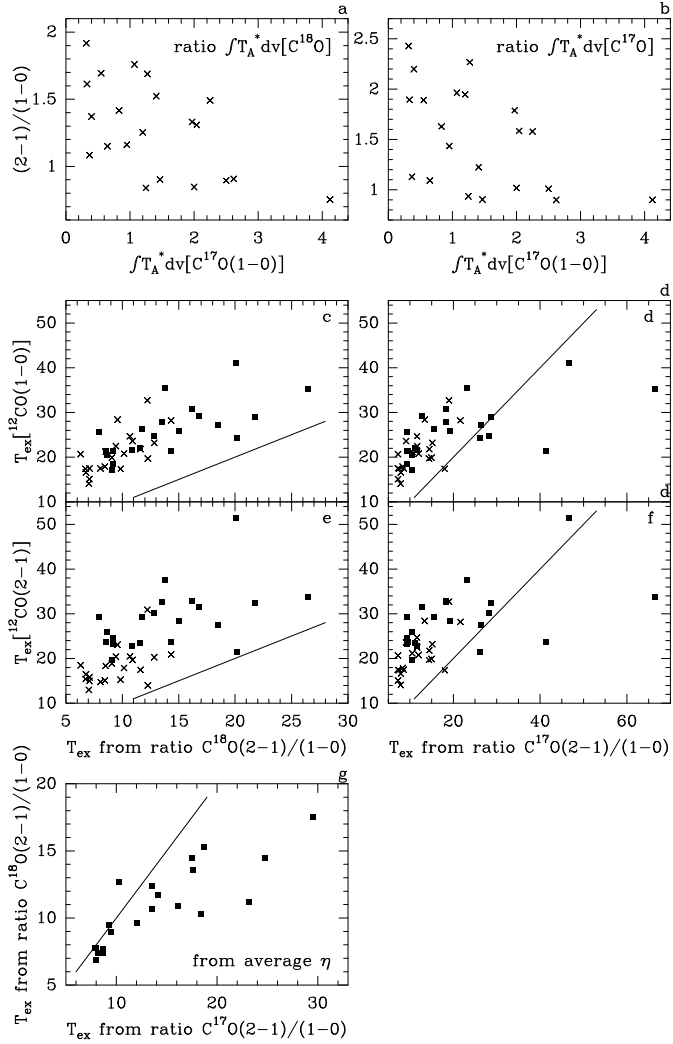}}
\caption{(
2--1)/(1--0) $\int$$T_{\rm A}^*$d$v$ 
ratios for C$^{18}$O ({\bf{a}}) and C$^{17}$O 
({\bf{b}}) as a function of $\int$$T_{\rm A}^*$[C$^{17}$O(1--0)]d$v$. 
{\bf{c,d}}. $T_{\rm {ex}}$[$^{12}$CO(1--0)] vs $T_{\rm {ex}}$ obtained from the ratio
(2--1)/(1--0) for C$^{18}$O ({\bf{c}}) and 
C$^{17}$O ({\bf{d}}) using $T_{\rm A}^*$ (crosses) or $T_{\rm mb}$ (squares). 
{\bf{e,f}}. The same as in c,d, but for  $T_{\rm {ex}}$[$^{12}$CO(2--1)].
In ({\bf{g}}) the $T_{\rm {ex}}$ from C$^{18}$O
and C$^{17}$O (using the average of Moon and main beam efficiencies) are compared. 
The drawn lines indicate equal values.
}
 \label{texcomp}
\end{figure}

Figs.\,\ref{texcomp}a, b show the (2--1)/(1--0) $\int$$T_{\rm A}^*$d$v$ ratios for C$^{18}$O 
and C$^{17}$O, respectively, as a function of $\int$$T_{\rm A}^*$[C$^{17}$O(1--0)]d$v$. 
There is no correlation.
In Figs.\,\ref{texcomp}c-f we compare $T_{\rm{ex}}$ derived from $J$=1--0 (Figs.\,\ref{texcomp}c,e)
and $J$=2--1 (Figs.\,\ref{texcomp}d,f)
$^{12}$CO $T_{\rm{A}}^*$ (Figs.\,\ref{texcomp}c,d)
values (this, method 1, should provide upper limits to $T_{\rm{ex}}$)
with the excitation temperatures of C$^{18}$O and C$^{17}$O, derived from the 
$\int$$T_{\rm A}^*$d$v$ (2--1)/(1--0) ratios (method 2). We also show in Figs.\,\ref{texcomp}e,f
the excitation temperatures after converting to 
$T_{\rm{mb}}$ using the efficiencies in Table~1 ($T_{\rm{mb}}$=$T_{\rm A}^*$$\eta_{moon}$/$\eta_{mb}$). 
The values derived from 
$T_{\rm A}^*$[$^{12}$CO] range from 14.1~K ($J$=1--0) and 13.0~K ($J$=2--1) at pos. 3 to 
32.7~K (1--0) and 30.9~K (2--1) at pos. 5. These values are for a $T_{\rm A}^*$ temperature scale. 
They are even higher for a $T_{\rm mb}^*$ scale: the maximum value is 41~K at pos. 5. 

Excitation temperatures derived from C$^{18}$O and C$^{17}$O (method 2) are generally lower: 6.3~K (pos. 
1) to 14.3~K (pos. 19) (C$^{18}$O) and 7.2~K (pos. 1) to 21.6~K (pos. 19) (C$^{17}$O). Using 
main beam brightness temperature ratios, the values become larger: 7.9 - 26.4~K (C$^{18}$O) 
and 9.3 - 66.4~K (C$^{17}$O) for the same positions. 

The OLS bisector mode was used to derive 
linear regression coefficients (see Isobe et al. \cite{isobe}). There is some correlation 
between the excitation temperatures derived from $^{12}$CO and C$^{18}$O (the correlation 
coefficient is 0.70 in Figs.\,\ref{texcomp}c). For C$^{17}$O, the correlation is slightly less 
well defined (the correlation coefficient is 0.66; Fig.\,\ref{texcomp}d). 
For C$^{18}$O the slope is closest to 1 for the $T_{\rm {mb}}$ ratios 
(0.93$\pm$0.16), while we obtain a slope of 1.97$\pm$0.33 for the $T_{\rm A}^*$ ratios. For 
C$^{17}$O this is reversed: for the $T_{\rm A}^*$ ratios the slope is 1.08$\pm$0.15 whereas 
that for the $T_{\rm {mb}}$ values it is 0.34$\pm$0.08. The higher values for $^{12}$CO 
compared to those from the C$^{18}$O and C$^{17}$O
(2--1)/(1--0) ratios can be explained by higher kinetic temperatures in the 
outer parts of the clouds (e.g. Castets et al. \cite{castets}) from which the $^{12}$CO emission
mostly originates. 

Because the extent of C$^{18}$O and C$^{17}$O clumps is most likely larger than the main beam, but
small compared to the size of the Moon we are using in Sects.\,4.2.2, 4.2.3, and 4.3 the average of 
both efficiencies. 
The resulting $T_{\rm {ex}}$ for C$^{18}$O and C$^{17}$O are compared with each other
in Fig.~\ref{texcomp}g. It shows
that $T_{\rm {ex}}$ is well correlated for both isotopomers, but for C$^{17}$O it can reach
higher values than for C$^{18}$O. There is no correlation between optical depth and $T_{\rm {ex}}$,
and therefore the lower $T_{\rm {ex}}$ of C$^{18}$O cannot be explained by the fact that
we did not correct for optically thick $J$=2$-$1 lines. 

\subsubsection{Optical depth}

C$^{18}$O optical depths are often derived from C$^{18}$O and C$^{17}$O data by assuming a 
certain intrinsic C$^{18}$O/C$^{17}$O ratio. However the aim of this paper is to determine 
this ratio and therefore this method cannot be applied.

We tried to fit the C$^{17}$O hyperfine components for several positions with small line widths. 
The optical depth for the main hyperfine component of the $J$=1--0 transition had values of 
less than a few tenths in most cases. The fitting is complicated by the presence of more than one 
velocity component, such as a broader underlying component (e.g. at pos. 10, which is also 
seen in C$^{18}$O), or two narrow 
components (pos. 3, 4, 7). In some cases we could fit line widths and velocities to the 
corresponding C$^{18}$O(1--0) spectrum and could use those as input values for the fit. 
Sometimes the fit gave a high optical depth for a weak component, which is not realistic. 
Limited signal-to-noise ratios prevented better determinations of the optical depth in these
components.  

We also derived optical depths from the excitation temperatures determined by method 2 using
the radiative transfer equation (see e.g. Rohlfs \& Wilson \cite{rohlfs}, Eq. (14.48)). We 
find that for excitation temperatures from (2--1)/(1--0) ratios using an average 
efficiency (as defined in Sect.\,4.2.1 and Fig.\,\ref{texcomp}g)
the highest 
total C$^{17}$O(1--0) optical depth (i.e. the resulting peak optical depth if all hyperfine
components had the same frequency) is $\tau_{\rm tot}$ = 1.09 for pos. 1. This is a little
high when considering $^{18}$O/$^{17}$O ratios of order 4 (Sect.\,5) and the results
of Bensch et al. (2001; see Sect.\,4.2). It is equivalent to a value for the 
main hyperfine component of about 0.36. At other positions $\tau_{\rm tot}$ ranges from 0.01 
to 0.43. For C$^{18}$O(1--0) the optical depth is undeterminable (log of negative number) at 
pos. 1, 3, 4 and 6 (but see Sect.\,4.2). At 
the other positions it ranges from 0.05 to 2.1. 

\subsubsection{LTE column density ratios}

Using the excitation temperature and optical depth derived above, the column density 
is calculated from

$$ N_{lte} = {3 h 10^5 1.065 \over 8 \pi^3 \mu^2} {\tau \Delta v \over J} Q
e^{E_{J-1} \over k T_{\rm {ex}}} \left[1 - e^{-h \nu /k T_{\rm {ex}}} \right]^{-1} $$

\hfill\break\noindent
with $\Delta v$ denoting the full width at half maximum intensity of the emission line, where the 
transition 
is from level $J$ to level $J-1$; $\nu$ is the observed frequency, $Q$ is the partition function (e.g. 
Rohlfs \& Wilson \cite{rohlfs}, Eq. (14.50)), and $\mu$ the electric dipole moment of the molecule.

Assuming that the structure of the cloud is somewhere in between that of a point source and a
very extended source, we used here for all following calculations telescope efficiencies which
are the average of the main beam and moon efficiency. 
Based on the results discussed in the previous sections we decided to adopt for C$^{18}$O and 
C$^{17}$O excitation 
temperatures and optical depths derived using the respective (2--1)/(1--0) ratios, and for 
$^{13}$CO the commonly adopted excitation temperature from the corresponding $^{12}$CO 
transition. In this way we derive
two sets of LTE column density ratios for the C$^{18}$O/C$^{17}$O and $^{13}$CO/C$^{18}$O ratios, 
one based on the $J$=1--0, the other on the 2--1 data.

The results for the 21 positions are given in Table~\ref{nratios}.
Col. 1 gives the position, 
Col. 2 the frequency-corrected C$^{18}$O/C$^{17}$O ratio, and Col. 3 the derived 
column 
density ratio. The C$^{17}$O excitation temperature, total optical depth and 
column density are in Cols. 4 to 6. At each position the first row is for $J$=1--0 and the
second row for $J$=2--1.
Cols. 7 and 8 give the excitation temperatures for 
C$^{18}$O and $^{12}$CO(1--0), respectively. 

\begin{figure}
 \resizebox{\hsize}{!}{\includegraphics{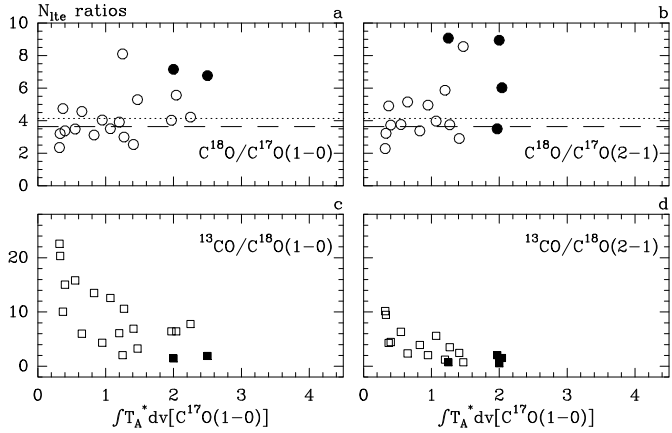}}
 \caption{C$^{18}$O/C$^{17}$O(({\bf {a, b}}) and $^{13}$CO/C$^{18}$O ({\bf {c, d}})
isotopomeric ratios as a function of $\int$$T_{\rm A}^*$[C$^{17}$O(1--0)]d$v$.
The ratios were derived from LTE column densities using
excitation temperatures derived from 
the corresponding (2--1)/(1--0) ratios, except for $^{13}$CO where we used $T_{\rm {ex}}$ 
derived from $^{12}$CO.
The filled symbols indicate positions where the C$^{17}$O 
excitation temperatures were used for the C$^{18}$O column density. 
The dashed line indicates the ratio found by Penzias (\cite{penzias}) for the galactic disk, 
and the dotted line marks the result from Bensch et al. (\cite{bensch}) from 
$^{13}$C$^{18}$O and  $^{13}$C$^{17}$O(1--0). }
 \label{ncoratios}
\end{figure}

The unweighted average $N$(C$^{18}$O)/$N$(C$^{17}$O) LTE ratio for the $J$=1--0 transition 
(see Fig.\,\ref{ncoratios}a) is
4.07$\pm$1.32 (sd; me 0.32), a higher value than that determined by Penzias (\cite{penzias}), 
but close to that derived by Bensch et al. (\cite{bensch}) from $^{13}$C$^{18}$O and 
$^{13}$C$^{17}$O. Omitting here the highest highly uncertain
value (above 8.0, at pos. 11) (the $\tau$ is close to being undetermined and therefore uncertain), the 
ratio becomes 3.81 $\pm$ 0.23). At four positions (1,3,4,6) 
the C$^{18}$O optical depth was undetermined, but for two of those positions (3,6) a value could
be derived using the C$^{17}$O excitation temperatures. However we did not use these 
data points to derive the average $N$(C$^{18}$O)/$N$(C$^{17}$O) ratio.
Similarly, for the $J$=2--1 transition using the same $T_{\rm{ex}}$ as above, the C$^{18}$O
optical depths were undetermined at positions 1, 2, 3, 4, 6, 7, 9, and 11, where at positions 1 to 4
also the C$^{17}$O $T_{\rm{ex}}$ resulted in undetermined $\tau$'s. Ratios derived for this
transition are shown in Fig.~\ref{ncoratios}b - the average value without the above mentioned positions
is 4.35$\pm$0.35. Omitting here the highest value (above 8.0, at pos. 8), 
the ratio becomes 4.00$\pm$0.29. 

$N$($^{13}$CO)/$N$(C$^{18}$O) ratios are shown in Fig.\,\ref{ncoratios}c,d, which were derived from
the $J$=1--0 and 2--1 data, respectively. 
One can see that 
after correction for optical depths the decrease in ratios towards the cloud 
center remains. This can be explained by real changes in the ratios such as fractionation; modelling 
them is beyond the scope of this paper. 

\begin{table*}
\caption[]{Derived parameters using average beam efficiencies.
\label{nratios}}
\begin{flushleft}
\begin{tabular}{rrr|rrr|rc|crcr}
\hline\noalign{\smallskip}
Pos & ratio &  & $T_{\rm ex}$ & $\tau$ & N & $T_{\rm ex}$ & $T_{\rm ex}$ & ratio & $T_{\rm kin}$ & log[n(H$_2$)]& $\chi^2$\\
    & \multicolumn{2}{c|}{C$^{18}$O/C$^{17}$O} &\multicolumn{3}{c|}{C$^{17}$O} & \multicolumn{1}{c}{C$^{18}$O} & \multicolumn{1}{c|}{$^{12}$CO(1--0)} & & & & \\
    & fcorr & N ratio & K &(tot) &cm$^{-2}$ & K & K  & C$^{18}$O/C$^{17}$O & K & cm$^{-3}$ &\\
    & \multicolumn{4}{c}{$J$=1$-$0}& & & & \multicolumn{4}{c}{LVG} \\
    & \multicolumn{4}{c}{$J$=2$-$1}& & & & \multicolumn{4}{c}{ } \\
\hline\noalign{\smallskip}
 1 &  2.18 &   -    &  8.0 & 1.09 & 7.2 10$^{15}$ &   6.9 &  22.9 &  4.5  & 9.4 & 5.19 & 4.5\\
   &  1.82 &   -    &      & 1.45 & 7.4 10$^{15}$ &       &       &       &     &      &    \\
 2 &  3.17 &   4.22 & 13.6 & 0.19 & 3.0 10$^{15}$ &  12.4 &  27.3 &  4.0  &17.8 & 4.89 & 6.8\\
   &  3.00 &   -    &      & 0.43 & 3.1 10$^{15}$ &       &       &       &     &      &    \\
 3 &  3.49 &   -    &  8.7 & 0.30 & 3.1 10$^{15}$ &   7.7 &  15.5 &  4.25 &11.5 & 4.71 & 18 \\
   &  3.09 &   -    &      & 0.47 & 3.0 10$^{15}$ &       &        &      &     &      &    \\
 4 &  3.01 &   -    &  7.9 & 0.43 & 3.4 10$^{15}$ &   7.8 &  19.3 &  3.75 &11.2 & 4.71 & 20 \\
   &  3.03 &   -    &      & 0.51 & 3.2 10$^{15}$ &       &       &       &     &      &    \\
 5 &  3.52 &   3.00 & 24.8 & 0.05 & 2.2 10$^{15}$ &  14.5 &  36.3 &  3.5  &23.2 & 4.71 & 11 \\ 
   &  2.62 &   3.76 &      & 0.15 & 2.3 10$^{15}$ &       &       &       &     &      &    \\
 6 &  3.67 &   -    &  8.7 & 0.28 & 2.4 10$^{15}$ &   7.4 &  18.3 &  5.0  &10.3 & 4.77 & 4.7 \\
   &  3.05 &   8.95 &      & 0.41 & 2.4 10$^{15}$ &       &       &       &     &      &    \\
 7 &  3.35 &   4.03 & 16.1 & 0.12 & 2.7 10$^{15}$ &  10.9 &  31.5 &  3.88 &15.4 & 4.80 & 2.7 \\
   &  2.49 &   3.51 &      & 0.33 & 2.9 10$^{15}$ &       &       &       &     &      &    \\
 8 &  3.72 &   5.30 &  8.0 & 0.19 & 1.7 10$^{15}$ &   7.8 &  16.6 &  4.5  &10.6 & 4.47 & 2.9 \\
   &  3.70 &   8.56 &      & 0.28 & 1.6 10$^{15}$ &       &       &       &     &      &    \\
 9 &  4.00 &   5.57 & 13.6 & 0.14 & 2.6 10$^{15}$ &  10.7 &  24.9 &  5.0  &14.8 & 4.83 & 0.9 \\
   &  3.30 &   6.03 &      & 0.29 & 2.7 10$^{15}$ &       &       &       &     &      &    \\
10 &  2.20 &   2.54 & 10.3 & 0.26 & 1.8 10$^{15}$ &  12.7 &  26.1 &  3.0  &12.2 & 4.59 & 14 \\
   &  2.75 &   2.91 &      & 0.38 & 1.7 10$^{15}$ &       &       &       &     &      &    \\
11 &  3.78 &   8.11 &  8.2 & 0.23 & 1.5 10$^{15}$ &   7.4 &  19.2 &  5.5  & 9.1 & 4.71 & 0.0\\
   &  3.40 &   9.08 &      & 0.36 & 1.4 10$^{15}$ &       &       &       &     &      &    \\
12 &  3.77 &   3.92 & 18.4 & 0.07 & 1.7 10$^{15}$ &  10.3 &  22.0 &  4.5  &10.3 & 4.71 & 14 \\
   &  2.43 &   5.88 &      & 0.17 & 1.8 10$^{15}$ &       &       &       &     &      &    \\
13 &  3.01 &   3.12 & 14.1 & 0.08 & 1.1 10$^{15}$ &  11.7 &  23.0 &  3.25 &11.2 & 4.41 & 0.7 \\
   &  2.64 &   3.38 &      & 0.14 & 1.0 10$^{15}$ &       &       &       &     &      &    \\
14 &  3.14 &   4.04 & 12.1 & 0.15 & 1.2 10$^{15}$ &   9.6 &  23.5 &  4.25 & 9.7 & 4.71 & 0.2 \\
   &  2.56 &   4.96 &      & 0.25 & 1.2 10$^{15}$ &       &       &       &     &      &    \\
15 &  3.63 &   3.51 & 18.7 & 0.05 & 1.6 10$^{15}$ &  15.3 &  25.7 &  3.88 &18.3 & 4.47 & 0.5 \\
   &  3.26 &   3.98 &      & 0.12 & 1.5 10$^{15}$ &       &       &       &     &      &    \\
16 &  3.52 &   3.21 & 17.6 & 0.03 & 4.6 10$^{14}$ &  13.6 &  24.1 &  3.5  &11.2 & 4.11 & 6.1 \\
   &  2.99 &   3.23 &      & 0.05 & 4.4 10$^{14}$ &       &       &       &     &      &    \\
17 &  3.82 &   4.57 &  9.3 & 0.12 & 7.4 10$^{14}$ &   9.5 &  19.7 &  4.63 &10.8 & 4.41 & 4.3 \\
   &  3.97 &   5.16 &      & 0.19 & 7.1 10$^{14}$ &       &       &       &     &      &    \\
18 &  3.72 &   3.49 & 17.5 & 0.03 & 7.6 10$^{14}$ &  14.5 &  21.7 &  3.75 &18.4 & 4.23 & 0.5 \\
   &  3.30 &   3.77 &      & 0.08 & 7.5 10$^{14}$ &       &       &       &     &      &    \\
19 &  3.22 &   2.36 & 29.5 & 0.01 & 6.1 10$^{14}$ &  17.5 &  31.3 &  3.0  &32.5 & 3.87 & 9.6 \\
   &  2.55 &   2.30 &      & 0.03 & 6.1 10$^{14}$ &       &       &       &     &      &    \\
20 &  4.39 &   4.75 &  9.5 & 0.04 & 4.0 10$^{14}$ &   9.0 &  19.3 &  4.5  &11.2 & 3.99 & 0.2 \\
   &  4.26 &   4.91 &      & 0.07 & 3.8 10$^{14}$ &       &       &       &     &      &    \\
21 &  4.36 &   3.38 & 23.2 & 0.02 & 6.6 10$^{14}$ &  11.2 &  19.1 &  4.25 &11.2 & 4.41 & 41 \\
   &  2.74 &   3.74 &      & 0.07 & 6.5 10$^{14}$ &       &       &      &      &    \\
\noalign{\smallskip}
\hline
\noalign{\smallskip}
\end{tabular}
\end{flushleft}
\end{table*}

\subsection{Isotopomeric ratios derived from Large Velocity Gradient (LVG) modelling}

In order to study the excitation of C$^{18}$O and C$^{17}$O in more detail, we also made LVG 
calculations (e.g. Castor \cite{castor}; Scoville \& Solomon \cite{scosol}). Simulating the 
observed line intensities, one can estimate
column densities, H$_2$ volume densities and kinetic temperatures. The critical point in our
LTE approach (Sect. 4.2) is the assumption of a single excitation temperature for all 
transitions of a given CO isotopomer. 
The LVG calculations provide a way to estimate
in how far differences in derived column densities
are accompanied by changes in excitation temperature. In principle, such changes could provide
line intensity ratios that do not directly reflect the $^{18}$O/$^{17}$O isotope ratio and an
LVG code is a suitable tool to investigate such effects. 

We used the collision rates from Flower (\cite{flower}) with an ortho/para H$_2$ ratio of 3.0.
Taking instead a ratio of 0.1 (i.e. almost pure para-H$_2$) does not significantly alter the 
results outlined below.
Level populations and expected line intensities were calculated for 
5~K $<$ $T_{\rm kin}$ $<$ 35~K and 10$^{3}$ $<$ $n$(H$_2$) $<$ 10$^{6}$~cm$^{-3}$.
For C$^{18}$O we assumed an average abundance of [C$^{18}$O]/[H$_2$] = 1.7~10$^{-7}$, derived 
by Frerking et al. (\cite{frerking}) for positions in the $\rho$ Oph cloud. 
In the 
calculations we used intrinsic ratios of C$^{18}$O/C$^{17}$O between 1.5 and 8.0 in steps of
0.25 or 0.5, and
adopted a velocity gradient of 5~km\,s$^{-1}$pc$^{-1}$, which is appropriate for the size and 
line width of the $\rho$ Oph clouds. 
 
Before discussing the results we have to check whether the assumptions used by Frerking et al. 
(\cite{frerking}) to derive C$^{18}$O column densities, $N$(C$^{18}$O),
are the same as used in the present paper. They observed 
C$^{18}$O(1--0) towards twelve positions and detected eight of them with a maximum $T_{\rm A}$ of
1.3~K. The positions were selected to have relatively low column densities in order to be able 
to derive extinctions ($A_{\rm v}$$<$10 mag, where depletion (see below) is not important).
Upper limits to column densities were derived assuming that the C$^{18}$O excitation 
temperature equals $T_{\rm {ex}}$ ($^{12}$CO) and that all levels are populated. To derive 
a lower limit to column densities Frerking et al. (\cite{frerking}) assumed the same excitation 
temperature, but with only the $J$=0 and 1 levels populated. The latter is clearly not true because 
we detected $J$=2--1 emission
in excess of 1.0~K at positions with $T_{\rm {A}}^*$[C$^{18}$O(1--0)] $<$ 1.0~K. Then Frerking 
et al. used the average value of upper and lower limit. 
Comparing
our LTE column densities (Sect.~4.2.3) with values derived using Frerking et al.'s assumptions 
shows
that Frerking et al. underestimate column densities by a factor of 2. Therefore the abundance is
too small by that amount.  

There is another reason why the assumption of a constant abundance is not necessarily correct. 
Kramer et al. (\cite{kramer}) 
found indications of CO depletion by a factor of about 3 in the core of \object{IC~5146} at visual 
extinction $A_{\rm v}$ of about 30 mag. IC~5146 is colder than the $\rho$ Oph cloud and 
Kramer et al. concluded that C$^{18}$O(1--0) and (2--1) are optically thin, which is not
the case for all positions in $\rho$ Oph. The depletion starts at visual
extinctions of about 10 mag (Bergin et al. \cite{bergin}), 
which would correspond to a $N$(C$^{18}$O) of $\approx$2 to 4 10$^{15}$ cm$^{-2}$ (assuming 
$N$(H$_2$)/$A_{\rm v}$ 0.9 10$^{21}$ cm$^{-2}$mag$^{-1}$; Bohlin et al. \cite{bohlin}). This is in 
the lower range of the column densities at the observed positions (see Table \ref{nratios}). 
However depletion is 
not expected to affect isotopomeric ratios and we have done some LVG calculations using a
C$^{18}$O abundance which is a factor of 2 lower than that of Frerking et al. (\cite{frerking}). 
The results differ only marginally from those obtained with an undepleted abundance
and are therefore inconclusive: at all positions the derived H$_2$ densities 
are approximately a factor of two larger and the kinetic temperatures are slightly 
smaller. 
Because the effects of depletion and the underestimation of column densities by Frerking 
et al. may balance out, we decided to use the Frerking et al. (\cite{frerking}) abundances in our 
calculations. 

Towards the 21 positions the observed and predicted values of peak C$^{18}$O temperatures (using 
averages efficiencies, but also $T_{\rm {A}}^*$ and $T_{\rm {mb}}$ for some calculations) 
and the C$^{18}$O/C$^{17}$O ratios for the $J$=1--0, 2--1 and 3--2 
lines were compared and $\chi$$^2$ values were calculated using for the temperatures an
uncertainty $\sigma$ of 10\% and for
the ratios of the three observed rotational transitions a $\sigma$ of 0.15, 0.15, and 0.20,
respectively. The measured temperatures for the $J$=2--1 and 3--2 transitions were convolved to
the $J$=1--0 beam, as described in Sect. 4.2.1.
Cols. 9 to 12 of Table \ref{nratios} give the results of the LVG 
calculations: C$^{18}$O/C$^{17}$O ratio, kinetic temperature, molecular hydrogen density
and $\chi^2$,
indicating the relative goodness of the solutions between the different positions. Apparently
the scatter in isotope ratio is smaller than that obtained with the LTE approach.
 
\begin{figure}
 \resizebox{\hsize}{!}{\includegraphics{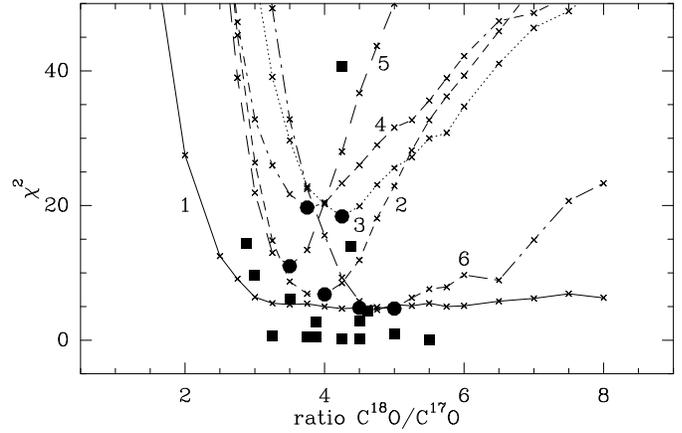}}
\caption{
Results of LVG calculations. Plotted is the derived $\chi$$^2$ value as a function
of the assumed C$^{18}$O/C$^{17}$O ratio for the positions 1 to 6. 
The different positions are distinguished 
by full drawn, dotted, and dashed lines. The filled symbols indicate the ratio of
minimum $\chi$$^2$ at each of the 21 positions. The circles are for the positions 1-6,
where $J$=3-2 data
exist. The positions associated with the filled squares can be identified in Col. 9 of 
Table~\ref{nratios}.
The small crosses mark the C$^{18}$O/C$^{17}$O ratio used in the calculations. 
}
 \label{chi2comp}
\end{figure}

\begin{figure}
 \resizebox{\hsize}{!}{\includegraphics{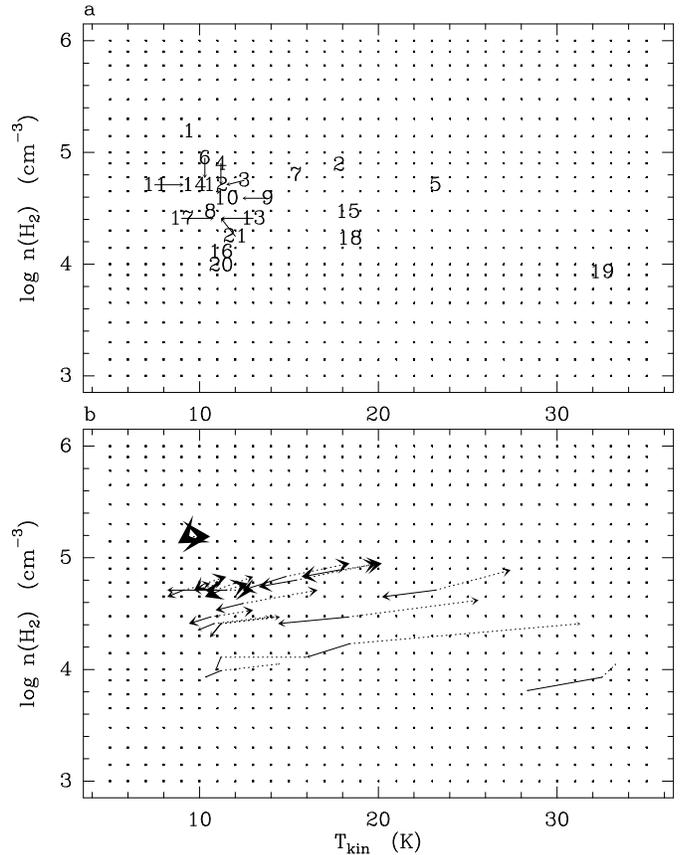}}
\caption{
{\bf{a}}. Results of LVG calculations. For each of the 21 positions
$T_{\rm kin}$ and log(n(H$_2$)) are given for the minimum $\chi$$^2$ displayed in
Fig.~\ref{chi2comp}. 
{\bf{b}}. The arrows show how $T_{kin}$ and log(n(H$_2$)) change if  
$T_{\rm {A}}^*$ (full-drawn arrows) and $T_{\rm {mb}}^*$ (dotted arrows) are used. The 
sizes of the arrowheads are proportional to $\int$$T_{\rm A}^*$[C$^{17}$O(1--0)]d$v$. 
In both panels the grid of dots indicate the $T_{\rm {kin}}$ and $n$(H$_2$) values 
used for the model calculations.
}
 \label{lvgresults}
\end{figure}

The parameters of minimum $\chi^2$ are shown in Fig.~\ref{chi2comp} (for the assumed
C$^{18}$O/C$^{17}$O ratio) and Fig.~\ref{lvgresults} (for $T_{\rm kin}$ and log(n(H$_2$))).
In Fig.~\ref{lvgresults}b the influence of the assumed temperature scale is also shown. 
The minimum $\chi^{2}$ values range between 0 (perfect fit) and more than 40 (bad fit; for pos. 21), 
and do not show a systematically lower value for either temperature scale. 
Fig.~\ref{lvgresults}b shows
a general decrease in density and an increase in $T_{\rm {kin}}$ with decreasing column density
(or $\int$$T_{\rm A}^*$[C$^{17}$O(1--0)]d$v$). LVG modelling assumes that the density is
constant along the line of sight, which is probably not true and some lines of sight may have
larger density gradients than others in the C$^{17}$O emitting region.
 
For all positions except of pos. 1, 
Fig.\ref{chi2comp} shows a clear minimum for some C$^{18}$O/C$^{17}$O.
Ratios ranges from 3.0 to 5.5 with an average value of
4.11$\pm$0.65 (sd; me 0.14). Giving a higher weight to positions with lower $\chi$$^2$ increases
the ratio slightly to 4.21. The exception is pos. 1 which has the highest H$_2$ density, where the 
$\chi$$^2$ does not much increase for higher ratios. 
We did not use $^{13}$CO data in these calculations because this isotopomer is too much 
affected by fractionation which makes it impossible to assume a single fractional abundance. Likewise
$^{12}$CO data show (much) higher excitation temperatures than the derived kinetic temperatures
(see also Fig.~\ref{texcomp}c-f).
This, like the increase in $T_{\rm kin}$ with decreasing column density mentioned above is
consistent with kinetic temperatures decreasing towards the cloud interiors from
about 20-30\,K to 10\,K. 

\begin{figure}
 \resizebox{\hsize}{!}{\includegraphics{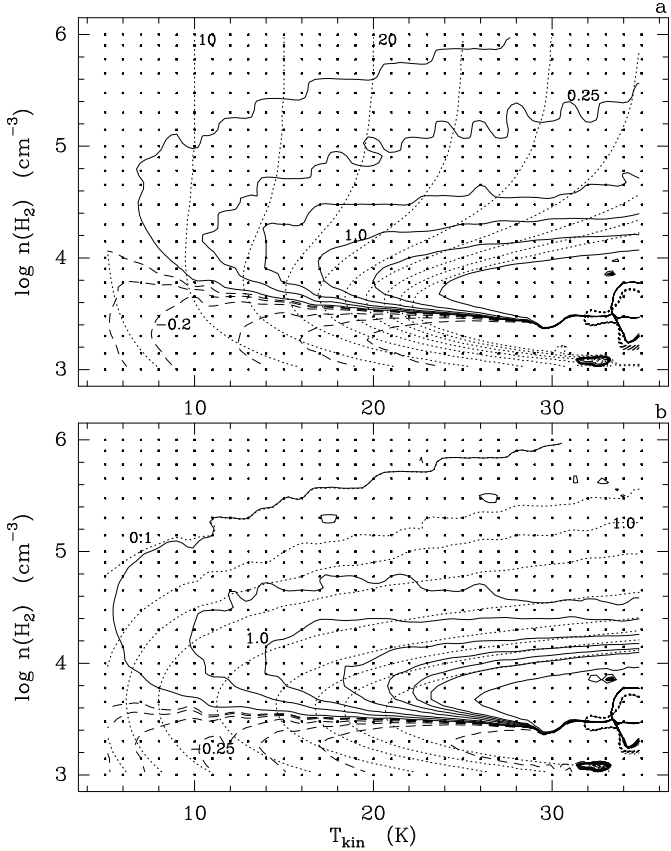}}
\caption{
A comparison of the $T_{\rm {ex}}$ of C$^{17}$O and C$^{18}$O derived from LVG model calculations. 
The grid of dots indicates the $T_{\rm {kin}}$ and 
$n$(H$_2$) values used for these calculations. 
{\bf{a}}.
The dotted lines indicate the non-LTE $T_{\rm {ex}}$ for C$^{17}$O(1--0) 
derived from the
LVG calculations using C$^{18}$O/C$^{17}$O=4.0. Contour levels are (left to right) 
5 to 50~K in steps of 5~K. The dashed and full-drawn lines show the difference in 
$T_{\rm {ex}}$ for the $J$=1$-$0 transition between C$^{17}$O and C$^{18}$O:
$T_{\rm {ex}}$(C$^{17}$O) -- $T_{\rm {ex}}$(C$^{18}$O).
Contour levels are -2, -1, -0.5, -0.2, -0.1, -0.05~K (dashed), 0.025, 0.25, 0.5, 1, 2, 5, 10~K. Some contour
values are labelled. 
{\bf{b}}.
The dotted lines show the difference in the non-LTE $T_{\rm {ex}}$ for two transitions of
C$^{17}$O: 
$T_{\rm {ex}}$[C$^{17}$O($J$=1$-$0)]--$T_{\rm {ex}}$[C$^{17}$O($J$=2$-$1)]. Contour levels are (left to right) 
0.1, 0.5, 1, 2, 5, 10, 15, 20, 30~K.
The dashed and full-drawn lines compare the differences in non-LTE $T_{\rm {ex}}$
for C$^{17}$O and C$^{18}$O: ($T_{\rm {ex}}$[C$^{17}$O($J$=1$-$0)]--$T_{\rm {ex}}$[C$^{17}$O($J$=2$-$1)]) --
($T_{\rm {ex}}$[C$^{18}$O($J$=1)]--$T_{\rm {ex}}$[C$^{18}$O($J$=2)]). Contour levels are -5, -1, -0.5, -0.25, -0.1, -0.05, -0.025 (dashed), 0.1, 
0.5, 1, 2, 3, 4, 5, 10~K. 
}
 \label{lvgtexcomp}
\end{figure}

We have to check what 
the LVG calculations predict for $T_{\rm {ex}}$, since LTE assumes that they are equal for all levels. 
Fig.~\ref{lvgtexcomp} shows results of the LVG calculations for an assumed
C$^{18}$O/C$^{17}$O ratio of 4.0. In Fig.~\ref{lvgtexcomp}a
the dotted lines indicate the predicted
$T_{\rm {ex}}$($J$=1$-$0) for C$^{17}$O. It is equal to $T_{\rm {kin}}$ at high densities, but for about
5 10$^3$ $<$ n(H$_2$) $<$ 5 10$^{4}$ cm$^{-3}$ $T_{\rm {ex}}$ is larger than $T_{\rm {kin}}$, while 
below 5 10$^3$ cm$^{-3}$ this situation is reversed. The dashed and full-drawn lines indicate the
difference in $T_{\rm {ex}}$($J$=1$-$0) between C$^{17}$O and C$^{18}$O. $T_{\rm {ex}}$(C$^{17}$O) is larger
than $T_{\rm {ex}}$(C$^{18}$O), in agreement with the observations (see Fig.~\ref{texcomp}g).
For $T_{\rm {kin}}$ $<$ 20~K the difference is less than 1~K. At low densities $T_{\rm {ex}}$(C$^{18}$O) 
is slightly larger than $T_{\rm {ex}}$(C$^{17}$O). 
In Fig.~\ref{lvgtexcomp}b the dotted lines show the difference in $T_{\rm {ex}}$ between the $J$=2$-$1 and $J$=1$-$0
transitions of C$^{17}$O. It is small for large n(H$_2$) or low $T_{\rm {kin}}$, but can reach values of more than
10~K for $T_{\rm {kin}}$ $>$ 20~K and n(H$_2$) about 10$^4$ cm$^{-3}$. However the difference of this
non-LTE effect between C$^{17}$O and C$^{18}$O (the full-drawn lines in Fig.~\ref{lvgtexcomp}b) is
much smaller: about 1~K or less in the region where most data points are located in Fig.~\ref{lvgresults}.
Also these results are insensitive to changes in the assumed ortho/para H$_2$ ratio: larger changes in the
excitation temperatures occur at higher temperatures ($T_{\rm {kin}}$$>$25~K) and lower densities 
($n$(H$_2$)$<$10$^4$~cm$^{-3}$) than those in the $\rho$ Oph region. 
This suggests that while LTE calculations will overestimate column densities both for C$^{17}$O and 
C$^{18}$O, the ratio of both column densities will be affected much less.  

\section{Discussion}

In calculating the isotope ratios we did not distinguish between the different velocity 
components in the cloud (with typical velocity differences of about 2~kms$^{-1}$).
The reason is the non-Gaussian line shape of many components 
which are confused by the hyperfine structure of C$^{17}$O (3 components for $J$=1--0 and 9 
components for $J$=2--1).  Assuming a depth along the line of sight of 0.4 pc (10\arcmin ), 
the crossing time for a velocity of 2~kms$^{-1}$ would be 2.0 10$^5$ yr, which is probably 
much less than the lifetime of the cloud. This is at least 5.5 10$^6$ yr, the age of the 
Upper Scorpius subgroup of the Scorpius-Centaurus OB association (de Geus et al. \cite{geus}).
Wilking et al. (\cite{wilkinglada}) obtained an upper limit for the ages of the T Tauri 
stars in $\rho$ Oph of 3.0 10$^6$ yr. This would provide enough time for sufficient mixing of 
the isotopic constituents of the gas. In addition the types of stars which produce $^{17}$O 
and $^{18}$O are not (yet) present in the $\rho$ Oph cloud (they do exist in the nearby Upper
Scorpius subgroup of the older Scorpius-Centaurus association), so both isotopes are not 
locally produced and it is unlikely that they are enhanced in some parts of the cloud by 
this mechanism (see e.g. Henkel \& Mauerberger \cite{henkelmau}). It also seems unlikely 
that stellar winds from these associations can alter the composition of the cloud significantly.

The results for the C$^{18}$O/C$^{17}$O ratios are summarized in Table~\ref{meanratios}, with 
the transition and the method used.

The first three entries in Table~\ref{meanratios} indicate that the observed
ratio is very dependent on the transition used. We note that
Penzias (\cite{penzias}) corrected his ratios only for the difference in
frequency and not for optical depth effects. This suggests that also the
$J$=1--0 ratios used in Penzias' (\cite{penzias}) galactic study need some more 
analysis. This will be discussed in more detail together with new measurements
in a forthcoming paper.
Within the uncertainties the average $J$=1--0 value is equal to the number obtained by Penzias 
(\cite{penzias}) for the galactic plane, 3.65$\pm$0.15.
The weighted average ratio of our LTE ratios is 4.17$\pm$0.26, whereas the LVG calculations
resulted in a ratio of 4.11$\pm$0.14. Because this method combines all observational data we
consider this the best result. 
Bensch et al. (\cite{bensch}) detected the almost certainly optically thin transitions 
$^{13}$C$^{18}$O and $^{13}$C$^{17}$O(1--0) towards pos. 1. Their observed 
$^{13}$C$^{18}$O/$^{13}$C$^{17}$O ratio, corrected for the frequency difference is 
4.23$\pm$0.53. Using escape probability models they derived a ratio $^{18}$O/$^{17}$O of 
4.15$\pm$0.52, which is identical to (but with a relatively large uncertainty because only one 
position was observed) the presently derived average value. Note that the position used by
Bensch et al. (\cite{bensch}) had to be omitted from our calculations because of an
undetermined optical depth. 

Recently Ladd (\cite{ladd}) observed C$^{18}$O and C$^{17}$O(1--0)
towards some 600 positions in the Taurus clouds and derives a ratio of 4.0$\pm$0.5, in 
agreement with the present result. However, the C$^{18}$O/C$^{17}$O ratio appears to decrease
with increasing integrated C$^{17}$O(1$-$0) intensity. Ladd then concludes that the ratio
in the inner parts is 2.8$\pm$0.4, due to larger self shielding of C$^{18}$O in the outer parts.
The range of integrated C$^{17}$O(1--0) intensity in Taurus (0.2 to 0.6 Kkms$^{-1}$) is much smaller
than in $\rho$~Oph because of the smaller line widths. We think that this ratio of 2.8 is not
real: our LVG models (using a smaller velocity gradient of 2.0 km\,s$^{-1}$\,pc$^{-1}$ and an
intrinsic C$^{18}$O/C$^{17}$O ratio of 4.0) can reproduce the decrease in ratio with $I$(C$^{17}$O)
if positions with lower $I$(C$^{17}$O) have a lower density and maybe kinetic temperature than
the points with higher $I$(C$^{17}$O), which is quite possible. In addition, it appears that when 
self-shielding is significant for C$^{18}$O, at $N$(H$_2$) $>$ 1.22 10$^{21}$ 
(Frerking et al. \cite{frerking}), or $A_{\rm v}$ $>$ 1.3 mag (Bohlin et al. \cite{bohlin}), this
corresponds to a 
far-UV extinction by dust of $>$13 mag (Aannestad \& Purcell \cite{aannestad}). This implies
that beyond this extinction (the (C$^{17}$O) self-shielding H$_2$ column density is even higher)
there is too little UV radiation left to affect the C$^{18}$O/C$^{17}$O ratio, in
contradiction with the suggestion by Ladd (\cite{ladd}).
In $\rho$~Oph a systematic decrease of C$^{18}$O/C$^{17}$O with $I$(C$^{17}$O)
is not seen (see Fig. \ref{coratios}), except towards pos. 1.

\begin{table}
\caption[]{Derived C$^{18}$O/C$^{17}$O ratios compared with some previous results. 
\label{meanratios}}
\begin{flushleft}
\begin{tabular}{lrll}
\hline\noalign{\smallskip}
Ratio & trans. & method & omit pos.\\
C$^{18}$O/C$^{17}$O & & \\
\hline\noalign{\smallskip}
3.53$\pm$0.11 & $J$=1$-$0 & freq. corr. & 1 \\
3.06$\pm$0.11 & $J$=2$-$1 & freq. corr. & 1 \\
2.78$\pm$0.18 & $J$=3$-$2 & freq. corr. & 1 \\
4.07$\pm$0.32 & $J$=1$-$0 & from $N_{\rm lte}$ & 1,3,4,6\\
4.35$\pm$0.44 & $J$=2$-$1 & from $N_{\rm lte}$ & 1,2,3,4,6,\\
              &           &        & 7,9,11 \\
4.17$\pm$0.26 & \multicolumn{3}{l}{average of $J$=1--0 and 2--1} \\
              &           &from $N_{\rm lte}$ & \\
4.11$\pm$0.14 & \multicolumn{3}{l}{$J$=1--0,2--1,3--2; from LVG}  \\
\noalign{\smallskip}
\hline
\noalign{\smallskip}
3.65$\pm$0.15 & $J$=1$-$0 & Penzias (\cite{penzias}) gal. disk & \\ 
              &           & freq.corr.& \\
4.15$\pm$0.52 & $J$=1$-$0 & Bensch et al. (\cite{bensch}) \\
  &           & from $^{13}$C$^{18}$O/$^{13}$C$^{17}$O \\
  &           & from exc. model\\
4.23$\pm$0.53 & $J$=1$-$0 & Bensch et al. (\cite{bensch}) \\
  &           & from $^{13}$C$^{18}$O/$^{13}$C$^{17}$O \\
  &           & freq. corr.\\
\noalign{\smallskip}
\hline
\noalign{\smallskip}
\end{tabular}
\end{flushleft}
\end{table}

\section{Summary}

From observations of up to three transitions of C$^{18}$O and C$^{17}$O towards 21 positions
in the $\rho$ Oph cloud we derive from LTE and LVG calculations
C$^{18}$O/C$^{17}$O abundance ratios of of 4.17 $\pm$ 0.26 and 4.11$\pm$0.1, respectively.
These are expected to be identical to the $^{18}$O/$^{17}$O isotope ratio. 
The average molecular hydrogen density towards the observed position 
increases from about 10$^4$ cm$^{-3}$ towards the postions with low column densities to 
10$^5$ cm$^{-3}$ towards positions in the cloud cores. The kinetic temperatures decrease from 30~K
or more at the edge of the cloud (as derived from the excitation temperatures of $^{12}$CO) to
20~K at positions with weak C$^{17}$O emission to 10~K in the cloud cores. 

\begin{acknowledgements}
This work was supported in part by the Deutsche Forschungsgemeinschaft 
through grant SFB-494.
The James Clerk Maxwell Telescope is operated by The Joint Astronomy Centre 
on behalf of the Particle Physics and Astronomy Research Council of
the United Kingdom, the Netherlands Organisation for Scientific Research, 
and the National Research Council of Canada. 
We thank Carsten Kramer for his comments on an earlier version of this paper. 
\end{acknowledgements}


\end{document}